\ifx\onecol\undefined
\documentclass[journal,twocolumn]{IEEEtran}
\else 
\documentclass[12pt,draftclsnofoot,journal,onecolumn]{IEEEtran}
\fi
\usepackage{amsfonts}
\usepackage{amsmath,amssymb}
\usepackage{acronym}  
\usepackage{algorithm}
\usepackage{algorithmic}
\usepackage{bm}
\usepackage{bbm}
\usepackage{booktabs}
\usepackage{color, soul}
\usepackage{cite}
\usepackage{graphicx}
\usepackage{indentfirst}
\usepackage{lipsum}
\usepackage{setspace}
\usepackage{tikz}
\usetikzlibrary{arrows}
\usepackage{subfigure}
\usepackage[amsmath,thmmarks]{ntheorem}
\usepackage{theorem}
\usepackage{hyperref}
\hypersetup{hidelinks, 
colorlinks=true,
allcolors=black,
pdfstartview=Fit,
breaklinks=true}

\usepackage{geometry}
\begin{document}

\title{Coded Beam Training}

\author{Tianyue Zheng, Jieao~Zhu, Qiumo~Yu, Yongli~Yan, and Linglong~Dai, {\textit{Fellow, IEEE}}
\thanks{
This work was supported in part by the National Key Research
and Development Program of China under Grant 2020YFB1807201, in part by 
by the National Natural Science Foundation of China under Grant 62031019, and in part by the European Commission through the H2020-MSCA-ITN META WIRELESS Research Project under Grant 956256.
{\it (Corresponding author: Linglong Dai.)}
}
\thanks{T. Zheng, J. Zhu, Q. Yu, Y. Yan, and L. Dai are with the Department of Electronic Engineering, Tsinghua University, Beijing 100084, China, and also with the Beijing National Research Center for Information Science and Technology (BNRist), Beijing 100084, China (e-mails: \{zhengty22, zja21, yqm22\}@mails.tsinghua.edu.cn, \{yanyongli, daill\}@tsinghua.edu.cn).
}}

\maketitle
\thispagestyle{empty}
\begin{abstract}
	In extremely large-scale multiple-input-multiple-output (XL-MIMO) systems for future sixth-generation (6G) communications, codebook-based beam training stands out as a promising technology to acquire channel state information (CSI). Despite their effectiveness, existing beam training methods suffer from significant achievable rate degradation for remote users with low signal-to-noise ratio (SNR). To tackle this challenge, leveraging the error-correcting capability of channel codes, we propose for the first time to incorporate channel coding theory into beam training to enhance the training accuracy, thereby extending the coverage area. Specifically, we establish the duality between hierarchical beam training and channel coding, and build on it to propose a general coded beam training  framework. Then, we present two specific implementations exemplified by coded beam training methods based on Hamming codes and convolutional codes, during which the beam encoding and decoding processes are refined respectively to better accommodate to the beam training problem. Simulation results have demonstrated that, the proposed coded beam training method can enable reliable beam training performance for remote users with low SNR, while keeping training overhead low.
\end{abstract}


\vspace{-1em}
\section{Introduction} \label{sec-intro}
Massive multiple-input multiple-output (mMIMO)~\cite{marzetta2010noncooperative} has been considered as a technological enabler for current fifth-generation (5G) communications. To achieve spectral efficiency enhancement in mMIMO systems, accurate channel state information (CSI) at the transmitter is a prerequisite, which can be realized either by the explicit CSI acquisition (i.e., channel estimation) or the implicit CSI acquisition (i.e., beam training)~\cite{9810144}. To further increase spectral efficiency, future sixth-generation (6G) communication systems are expected to employ extremely large-scale MIMO (XL-MIMO) antenna arrays~\cite{6G,10379539}. Unfortunately, due to the high-dimensional XL-MIMO channels, the pilot overhead of channel estimation will increase dramatically, making explicit CSI acquisition challenging~\cite{9452036,9508929}. In such cases, implicit CSI acquisition, (beam training), serves as a more important and practical way to acquire the CSI \cite{9913211}. This implicit procedure is performed by transmitting several predefined directional beams (codewords) towards the users, and determining the users' direction from their received power~\cite{10130629,7438800}.

An important way to conduct beam training is exhaustive beam sweeping~\cite{exhaustive,7845674,MUexhaus}, which sequentially tests the narrow beams predefined in the codebook, and selects the codeword with the highest received power. Thanks to the high beamforming gain of narrow beams, exhaustive beam sweeping by narrow beams ensures reliable beam training performance even for remote users (usually located in the cell-edge area) with low signal-to-noise ratio (SNR)~\cite{exhaustive}. Despite the considerable reliability of exhaustive beam sweeping, the size of the exhaustive codebook grows linearly with the number of BS antennas~\cite{7845674}. Thereby, the exhaustive beam sweeping brings unaffordable training overhead in XL-MIMO communication systems.  

To reduce the training overhead, hierarchical beam training methods have been proposed \cite{tdma,9547829,10057262,10239282}. In hierarchical beam training, the possible user directions are narrowed down in a layer-by-layer manner. Specifically,  hierarchical beam training utilizes a hierarchical codebook comprising multi-layer codewords. In this codebook, the spatial region covered by a certain codeword at any layer of the codebook is partitioned into two smaller disjoint spatial regions in the next layer~\cite{tdma}. 
Then, applying this codebook, the BS can  gradually narrow down the possible user direction by choosing the spatial region with larger received power based the user's feedback signal in each layer.
Owing to the half reduction of uncertain region of the user's direction in each layer, this hierarchical scheme can exponentially speedup the beam training process compared with the exhaustive beam sweeping, contributing to a remarkable enhancement of the spectral efficiency~\cite{9547829}. Thus, the idea of hierarchical beam training has triggered various improvement efforts for designing binary search-based hierarchical codebooks~\cite{deact,tdma,channel,sum,9547829,10057262}. However, due to the ``error propagation'' phenomenon, hierarchical beam training methods suffer from serious performance deterioration for remote users with low SNR. The error propagation originates from low directional beam gain of wide beams in the upper layers. With reduced beam gain, these upper-layer beams are especially susceptible to errors, causing unrecoverable errors in the subsequent layers of the hierarchical process.  

To our best knowledge, existing beam training methods, both exhaustive and hierarchical, can hardly resolve the conflict of reliability and efficiency in beam training for remote users with low SNR. To fill in this gap, in this paper, we propose a coded beam training framework by introducing channel coding into beam training. Exploiting the error correction capability of channel codes, the proposed  framework enables reliable beam training performance with exponentially reduced pilot overhead, even for the remote users. Specifically, the main contributions of this paper are summarized as follows\footnote{Simulation codes will be provided to reproduce the results in this paper: \url{http://oa.ee.tsinghua.edu.cn/dailinglong/publications/publications.html}.}.
\begin{enumerate}
	\item By analyzing of the binary algebraic structure of hierarchical beam training, this paper is the first attempt to reveal the duality of hierarchical beam training problem and channel coding problem, based on which a unified coded beam training framework is proposed. Leveraging this duality, almost all kinds of channel codes can be seamlessly integrated into the proposed coded beam training framework. 
	\item To perform coded beam training, we design the space-time coded beam patterns for generating the codewords and the transmitting beamformers during beam training, where different spatial directions are encoded into different time sequences based on the channel encoder to improve the tolerance to noise. Then, we utilize the sequence of received signal power to decode the spatial directions of the user by exploiting the error correction capability of channel codes, yielding the desired codeword for the user.  
	\item To better accommodate to the beam training problem, we improve the coding algorithms in two aspects. Firstly, existing channel coding algorithms are designed for Gaussian channel for data transmission, while the user's received power during beam training obeys a non-central $\chi^2$ distribution. Therefore, we modify the log-likelihood ratio (LLR) calculator in the beam training decoder to better adapt to the probabilistic properties in the beam training problem. Secondly, we propose an adaptive encoding process where we dynamically adjust the coded beam pattern based on the outcomes of previous decoding iterations. The adaptive beam training encoder can exclude impossible directions, thus improve the real-time beam gain. 
	\item We employ classical Hamming codes and convolutional codes respectively as examples to illustrate our proposed coded beam training. We provide simulation comparison of our proposed coded beam training method with other methods, demonstrating the proposed coded beam training method can enable reliable beam training for remote users with low SNR. Besides, simulation results also validate that the $\chi^2$ decoder outperforms the traditional Gaussian decoder.
\end{enumerate}

The rest of this paper is organized as follows. 
In Section~\ref{Sec_3}, the system as well as channel models are introduced, and the problem of beam training is formulated. Then, the principles and implementation of the proposed coded beam training are elaborated in Sections~\ref{CBT} and \ref{Implementation}, respectively. Simulation results are provided in Section~\ref{sec-re}. Finally, Section~\ref{sec-con} concludes this paper.

{\it Notations}: Lower-case and upper-case boldface letters represent vectors and matrices, respectively; $[N]$ denotes the set $\{0, 1, \cdots, N-1\}$; $\| \cdot \|_p$ denotes the $p$-norm of a vector; $\mathbb{C},\mathbb{R}$ denote the set of complex and real numbers, respectively; $[\cdot]^T, [\cdot]^H$ denote the transpose, and conjugate-transpose operations, respectively; $\bigcup, \bigcap$ denote the union and intersection operation of sets; $\mathcal{C} \mathcal{N}(\mu,{\rm \varSigma })$ denotes the Gaussian distribution with mean $\mu$ and covariance ${\rm \varSigma }$;  $\oplus$ denotes the exclusive OR (XOR) operation; $I_\nu$ denotes the $\nu$-th order modified Bessel function of the first kind.

\section{System Model} \label{Sec_3}
In this section, the channel model of the XL-MIMO system used in this paper will be introduced first. Then, we will formulate the beam training problem.

\subsection{System Model} \label{Sec_3_Subsec_1} 
We consider a mmWave/Terahertz (THz) XL-MIMO system with one base station (BS) and a single-antenna user equipment (UE). The BS employs a uniform linear array (ULA) equipped with $N_{\rm T}$ $\lambda/2$-spaced antennas, each being connected to one RF chain, i.e., we adopt the fully-digital precoding structure. It is worth emphasizing that our main technical contributions are not restricted to full-digital precoding and can be extended to arbitrary precoding architecture by applying corresponding beam design methods ~\cite{tdma,8408778,Hybrid,9411813,9618146,10294206}, examples include the hybrid precoding elaborated in Section~\ref{Sec_4_Subsec_4}.

For the downlink transmission, let $s_0 \in \mathbb{C}$ be the power-normalized transmitted symbol, then the received signal $y$ at the UE is given by 
\begin{equation} \label{eq-signal}
	y = \sqrt{P} {\bm h}  {\bm w} s_0 + n,
\end{equation}
where $P>0$ is the transmit power, ${\bm h} \in \mathbb{C}^{1 \times N_{\rm T}}$ the downlink channel, ${\bm w} \in \mathbb{C}^{N_{\rm T} \times 1}$ the unit-norm transmit beamformer,  and $n$ the complex circularly-symmetric additive white Gaussian noise $n \sim \mathcal{CN}(0,\sigma^2)$ at the UE receiver. 

According to the well-known Saleh-Valenzuela channel model~\cite{channel}, the channel ${\bm h}$ can be expressed as
\begin{equation}\label{eq-channel}
	{\bm h} = \sqrt{\frac{N_{\rm T}}{L_0}} \sum_{l = 1}^{L_0} \beta_l {\bm \alpha}(\varphi_l),
\end{equation}
where $L_0$ is the number of multipath components, $\beta_l$ and $\varphi_l$ represent the complex gain and the angle-of-departure (AoD) of the $l$-th path. ${\bm \alpha}(\varphi) \in \mathbb{C}^{1 \times N_{\rm T}}$ is the array steering vector, which is defined as 
\begin{equation}\label{eq:steervector}
	{\bm \alpha}(\varphi)=\frac{1}{\sqrt{N_{\rm T}}} [1,e^{-j\pi \varphi,},\ldots,e^{-j(N_{\rm T}-1) \pi \varphi}],
\end{equation}
where $\varphi \triangleq \sin \theta \in [-1,1]$ denotes the spatial direction and $\theta\in [-\pi/2, \pi/2]$ the physical direction. 
The significant scattering attenuation at high frequencies makes the power of the LoS path considerably higher than its NLoS counterparts, rendering the former a dominant component for data transmission at mmWave/THz bands. Therefore, this paper mainly considers the channel with only LoS component, which also determines the pointing direction from the BS to the UE~\cite{9508929}.

\subsection{Problem Description} \label{Sec_3_Subsec_2}
The objective of beam training is to steer the beamformer ${\bm w}$ to the AoD of the dominate path (LoS path). Specifically, according to the structure of the array steering vector in \eqref{eq:steervector}, we define the DFT codebook, $\mathcal{W}$, also known as the exhaustive codebook, as
	\begin{equation} \label{bottom}
		\begin{aligned}
			\mathcal{W} = \{ {\bm \alpha}(\varphi) \vert  \varphi = -1+(2n-1)/N_{\rm T},n \in \{1,2,\cdots, N_{\rm T}\} \}.
		\end{aligned}
	\end{equation}
	the diagram of which is illustrated in Fig.~\ref{fig:codebook}(a). 
Codebook-based beam training promises to select a codeword from $\mathcal{W}$ to maximize the received signal power, i.e., 
\begin{equation} \label{prob_sol}
	\begin{aligned}
		&\max \limits_{{\bm w}} \quad \vert {\bm h}  {\bm w} \vert  \\
		&{\rm s.t.} \quad  {\bm w} \in \mathcal{W}. \\
	\end{aligned}
\end{equation}

To solve the problem~\eqref{prob_sol}, a straightforward way is to perform exhaustive beam sweeping~\cite{exhaustive}. The BS first sequentially sweeps all codewords from  $\mathcal{W}$. Then, the UE selects the best codeword having the highest received signal power and feedbacks the selected codeword index. Clearly, the beam sweeping process occupies $N_{\rm T}$ time slots, equivalent to the number of the BS antennas. This fact means that although this exhaustive beam sweeping could achieve a good beam training performance, it inevitably consumes unaffordable training overhead, especially for XL-MIMO systems. 

\ifx\onecol\undefined
\begin{figure}
	\centering 
	\subfigure[Exhaustive codebook.]{
		\includegraphics[width=0.95\linewidth]{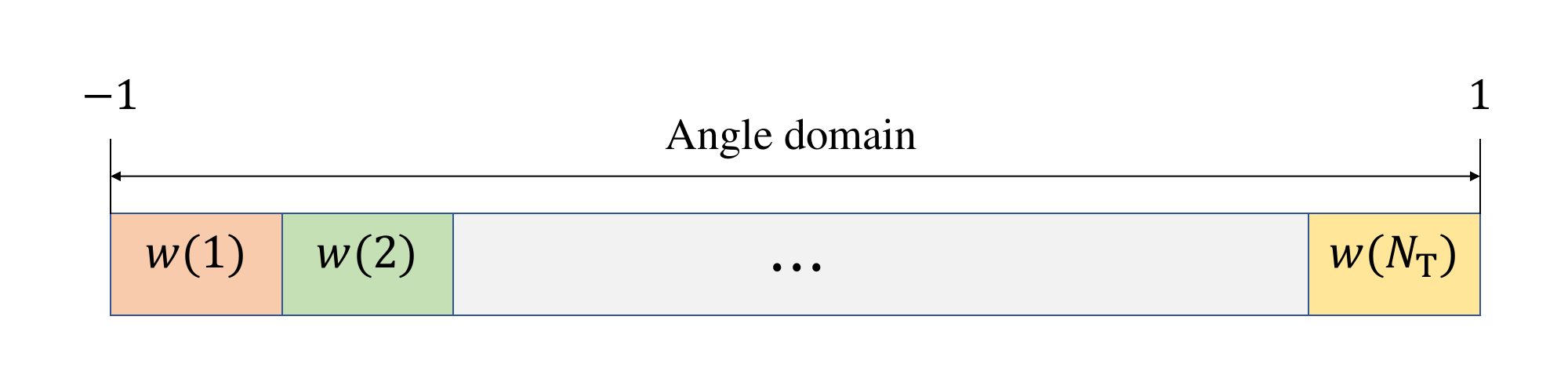}} 
	\subfigure[Hierarchical codebook.]{
		\includegraphics[width=0.95\linewidth]{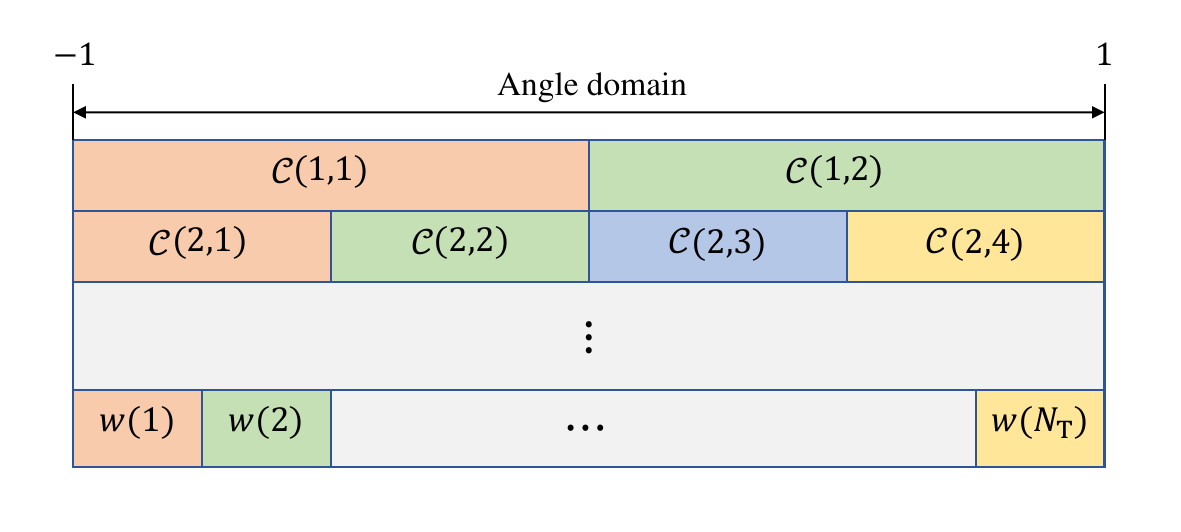}}
	\caption{Illustration of the DFT codebook, $\mathcal{W}$, and the hierarchical codebook, $\mathcal{C}_{\rm hier}$, where the subscript ``hier" in omitted in the figure.}
	\label{fig:codebook}
\end{figure}
\else 
\begin{figure}
	\centering 
	\subfigure[Exhaustive codebook.]{
		\includegraphics[width=0.65\linewidth]{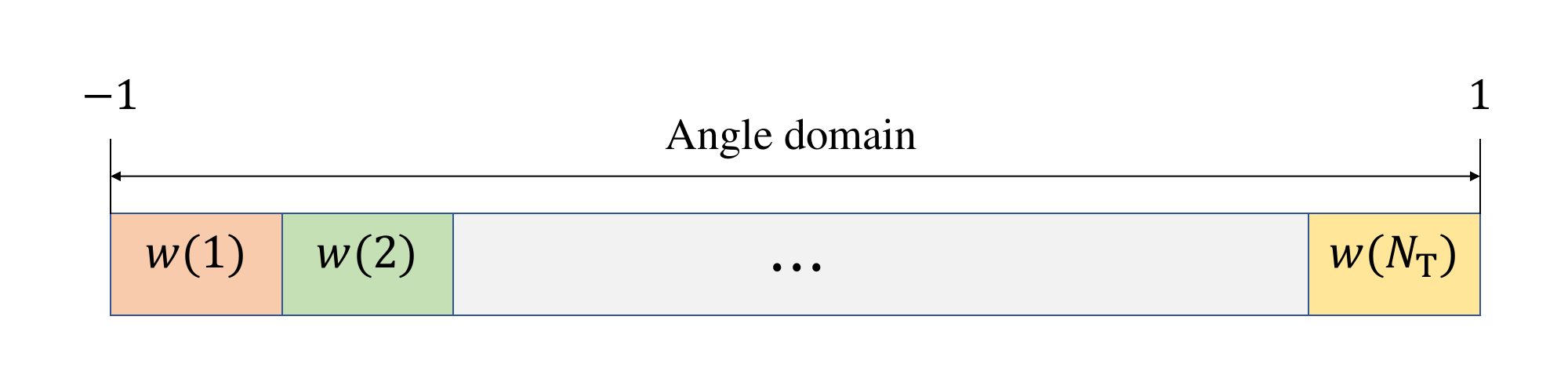}} 
	\subfigure[Hierarchical codebook.]{
		\includegraphics[width=0.65\linewidth]{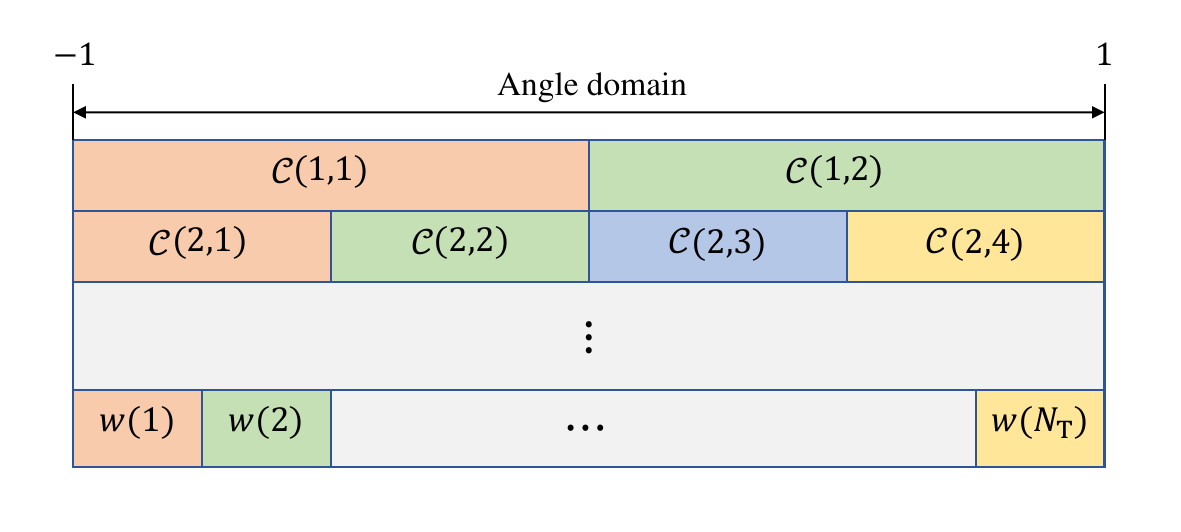}}
	\caption{Illustration of the DFT codebook, $\mathcal{W}$, and the hierarchical codebook, $\mathcal{C}_{\rm hier}$, where the subscript ``hier" in omitted in the figure.}
	\label{fig:codebook}
\end{figure}
\fi

To avoid the unacceptable training overhead incurred by exhaustive beam sweeping, hierarchical beam training utilizing binary-search based codebook are widely adopted.
As presented in Fig.~\ref{fig:codebook}(b), a typical hierarchical codebook~\cite{tdma}, {$\mathcal{C}_{\rm hier}$}, has $2^l$ codewords at the $l$-th ($l\in \{1,2,\cdots,\log_2 N_{\rm T}$\}) layer,   
each of which, denoted as $\mathcal{C}_{\rm hier}(l,b)$, covers two higher-resolution codewords with narrower coverage angle at the $l+1$-th layer.
During the beam training process, we test the power of the received signal with two selective low-resolution codewords at the upper layer, choose the one with higher received power,
and then narrow down the beam width in a layer-by-layer manner, until a specific codeword at the bottom layer is obtained. Through this way, the beam training overhead is reduced to $2 \log_2 N_{\rm T}$~\cite{tdma}. However, the performance of hierarchical beam training suffers from the ``error propagation'' phenomenon,  and thus cannot ensure reliable beam training for remote users with low SNR, leading to a restricted coverage area. Specifically, the codewords at higher layers have wider beamwidth and lower beam gain, making it more vulnerable to noise.
Since hierarchical beam training works on a binary tree in a sequential manner, any erroneous decision at some layer on the tree will lead to unrecoverable training failure.


In this paper, to alleviate the ``error propagation'' curse, we propose a new hierarchical beam training method utilizing the self-correcting capabilities of channel codes, which can reduce training overhead while maintaining the beam training success rate for remote users with low SNR.  

\section{Overview of Coded Beam Training} \label{CBT}

Channel codes are well-established error control techniques to protect the transmission data against channel noise by adding redundant bits. Compared to non-coded systems, coded systems are able to dramatically decrease the bit error rate (BER) under the same channel condition and data payload requirements, which is known as the waterfall effect of the BER. 
Inspired by channel coding, we develop an ultra-reliable hierarchical beam training framework, namely {\it coded beam training}. In the proposed framework, exploiting the error correction capability, channel codes are introduced to hierarchical beam training by adding extra layers of codewords to protect the hierarchical beam training process from channel noise. 

This section elaborates on the principles of coded beam training. We first illustrate the fundamental idea of coded beam training using an introductory example of a $(7,4)$ Hamming code. Then, this idea is extended to a general framework of coded beam training.


\subsection{An Introductory $(7,4)$ Hamming Code Example}\label{ham}
To help the understanding of the proposed framework, we start from comparing the traditional binary-search based hierarchical beam training~\cite{tdma} with the coded beam training exploiting $(7,4)$ Hamming code in an $N_{\rm T} = 16$-antenna system.

\subsubsection{Motivation of Coded Beam Training} In traditional hierarchical beam training, the codebook $\mathcal{C}_{\rm hier}$ contains $L=\log_2 N_{\rm T}=4$ layers, the $l$-th layer of which has $2^l$ codewords. The detailed beam training procedure is carried out as follows. 
We divide the spatial direction into $N_{\rm T}=16$ segments uniformly and let the length-4 bit ${\bf u}\in \{0, 1\}^4$ label the spatial direction of the user. At the first layer of codebook $\mathcal{C}_{\rm hier}$, the BS sequentially transmits $\mathcal{C}_{\rm hier}(1,1)$ and $\mathcal{C}_{\rm hier}(1,2)$ to the UE. 
Then, the UE compares the received signal power of $\mathcal{C}_{\rm hier}(1,1)$ and $\mathcal{C}_{\rm hier}(1,2)$, and set ${\bf u}(1)=0$ if the first codeword yields higher signal power, and ${\bf u}(1)=1$ otherwise. 
After that, the UE feeds back the bit ${\bf u}(1)$ to the BS, according to which the BS selects the two possible codewords in the second layer. We sequentially perform these steps until approaching the bottom layer of $\mathcal{C}_{\rm hier}$. 
The BS can finally decide the optimal index according to the bitstream ${\bf u}=[{\bf u}(1),{\bf u}(2),{\bf u}(3),{\bf u}(4)]$. 
For example, if the received bitstream ${\bf u}= [0,0,1,0]$, the selected codeword index is $ {\rm bintodec}([0,0,1,0])+1=3$. According to (\ref{bottom}), beamformer ${\bm w} = {\bm \alpha}( -1 + (2\times3 - 1)/16) = {\bm \alpha}( -11/16)$ is adopted for data transmission. However, for a remote user with low SNR, if the first layer of codebook chooses the wrong index due to low directional gain of wide beam, the subsequent training process is invalid since we have missed the optimal index. 
Suppose the original ${\bf u}(1)=0$ is incorrectly decided as ${\bf u}(1)=1$, then the selected codeword index is $11$ instead of $3$. This issue is referred to the ``error propagation'' problem, which will be alleviated using the Hamming code to protect the index against incorrect bits. 

\begin{figure*}
	\centering 
	\includegraphics[width=\linewidth]{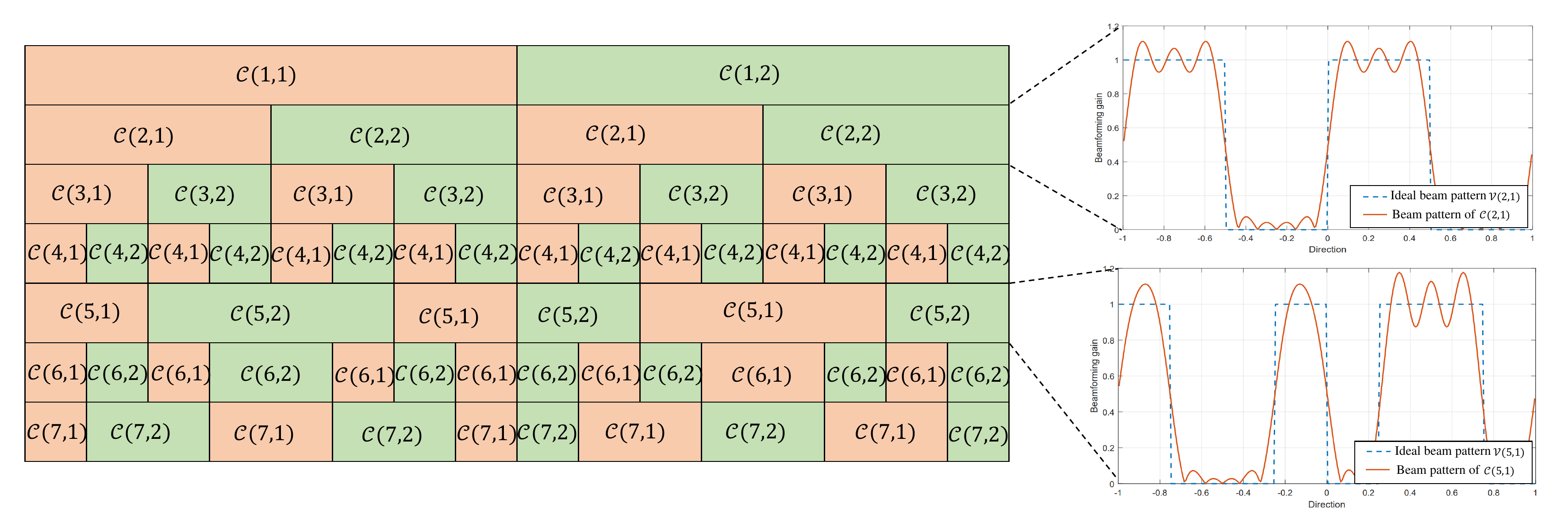}
	\caption{The codebook $\mathcal{C}_{\rm ham}$ designed by the (7,4) Hamming code (14 codewords in 7 layers in total), where the subscript is omitted in the figure. }
	\label{fig:hamming}
\end{figure*}

\subsubsection{Beam Encoding} 
To alleviate the error propagation issue, we consider the incorporation of $(7,4)$ Hamming code and beam training. Our novelty lies in the design of a new codebook, {say $\mathcal{C}_{\rm ham}$}, comprising seven layers (four information layers and \emph{three additional check layers}) as presented in Fig. \ref{fig:hamming}. 
Compared with the classic hierarchical codebook, $\mathcal{C}_{\rm hier}$, the proposed codebook,  $\mathcal{C}_{\rm ham}$, contains two complementary codewords in each layer, featuring  sawtooth-shaped beam patterns, which make it capable of encoding beams. 
The detailed construction of $\mathcal{C}_{\rm ham}$ involves two step: 1) beam pattern design relying on the Hamming code, 2) codebook generation based on the beam pattern, which are elaborated below.

\begin{itemize}
	\item \textbf{Beam pattern design}: 
	To begin with, it is necessary to determine the space-time $0$-$1$ beam pattern, $\mathcal{V}_{\rm ham}$, for building the codebook $\mathcal{C}_{\rm ham}$. Each element of $\mathcal{V}_{\rm ham}$ reflects the desired beam pattern of one codeword of $\mathcal{C}_{\rm ham}$. Specifically, for the desired beam generated by the $b$-th codeword in $l$-th layer, $\mathcal{C}_{\rm ham}(l,b)$, its beam pattern $\mathcal{V}_{\rm ham}(l,b)$ is composed of $N_{\rm T} = 16$ binary numbers, i.e., $\mathcal{V}_{\rm ham}(l,b) = \{0,1\}^{16}$.  
	By dividing the spatial region $[-1,1]$ into $N_{\rm T}=16$ segments uniformly, $\mathcal{V}_{\rm ham}(l,b,i)$ is set as $0$ if the beam generated by codeword $\mathcal{C}_{\rm ham}(l,b)$ is expected to have a high gain in the $i$-th segment ($i \in [16]$) and $\mathcal{V}_{\rm ham}(l,b,i)=1$ otherwise. 
	
	Applying this definition, we can build $\mathcal{V}_{\rm ham}$ using the Hamming codes. To be specific, we first index the $i$-th spatial segment, $i\in[16]$, by a length-$k=4$ bits ${\bf u}_i$. {According to the encoding operation of $(7,4)$ Hamming coding, the coded bitstream ${\bm x}$ is expressed as }
	\begin{equation}
		{\bm x}_i= {\bf u}_i {\bf G} \in \{0,1\}^{7},
	\end{equation}
	where $ {\bf G}$ is the generator matrix, which is denoted as 
	
	\begin{equation}
		{\bf G}=\left[ \begin{array}{cccc|ccc}
			1 & 0 & 0 & 0 & 1 & 1 & 1 \\
			0 & 1 & 0 & 0 & 1 & 1 & 0 \\
			0 & 0 & 1 & 0 & 1 & 0 & 1 \\
			0 & 0 & 0 & 1 & 0 & 1 & 1
		\end{array}\right].
	\end{equation}
	In our codebook design, the $l$-th entry of ${\bm x}_i$ defines the beam pattern of the codeword $\mathcal{C}_{\rm ham}(l,1)$ over the $i$-th segment, while its flip determines the beam pattern of $\mathcal{C}_{\rm ham}(l,2)$.
	Therefore, the corresponding space-time beam pattern $\mathcal{V}_{\rm ham}(l,b), l\in\{1,2,\cdots,7\},b\in\{1,2\}$ can be expressed as
	\begin{equation}\label{eq:74}
		\left\{\begin{array}{cc} \mathcal{V}_{\rm ham}(l,1,i) &= {\bm x}_i(l)\\
			\mathcal{V}_{\rm ham}(l,2,i) &= \bar{\bm x}_i(l) 
		\end{array} \right. , \quad  i \in [16], 
	\end{equation}
	where $\bar{x}$ denotes the bit flip and ${\bm x}(i)$ denotes the $i$-th bit of a vector ${\bm x}$. The designed beam pattern by \eqref{eq:74} is illustrated in the left part of Fig.~\ref{fig:hamming}.
	
	\item \textbf{Codebook Generation}: After obtaining the space-time beam pattern, we are able to generate the codebook, $\mathcal{C}_{\rm ham}$, employing various beam design methods, such as the
	weighted sum of narrow beams~\cite{beamgene} and the Gerchberg-Saxton (GS) algorithm~\cite{10365224}.
\end{itemize}

As illustrated in Fig.~\ref{fig:hamming}, the first four layers of the beams are regular square beams and the extra three check layers are irregular multi-mainlobe wide beams.
In each layer, the BS sequentially sends all codewords to the UEs. The UEs compare the received signal power of two codewords and feedback one bit to label the stronger codeword. In the same example, if the original spatial information bitstream is ${\bf u}= [0,0,1,0]$, then the desired feedback bit time sequence is ${\bm \hat{x}}= [0,0,1,0,1,0,1]$.

\subsubsection{Beam Decoding}
In the beam decoding part, we aim to decide the optimal codeword (i.e. space information) index according to the received bitstream. If we suppose the first layer is decided incorrectly due to the influence of noise again in this example, the received bitstream will change to ${\bm \hat{x}}= [1,0,1,0,1,0,1]$. Then we will illustrate how we obtain the correct index with the error correction ability of Hamming code.

Based on the parity check matrix ${\bf H}$, Hamming decoder helps determine whether the received bitstream contains error bit and the specific position of the error bit. The syndrome is computed as 
\begin{equation}
	{\bf c}={\bm \hat{x}} {\bf H^T},
\end{equation}
where parity check matrix can be expressed as 
\begin{equation}
	{\bf H}=\left[\begin{array}{cccc|ccc}
		1&1&1&0 &1&0&0 \\
		1&1&0&1 &0&1&0 \\
		1&0&1&1 &0&0&1
		\end{array}\right].
		\label{equ:check} 
\end{equation}
Then we can decide the error pattern based on the syndrome as in Table.~\ref{check}.

\vspace{1em}
\begin{table} 
	\centering
	\caption{Error patterns and the corresponding syndromes of $(7,4)$ 
	Hamming code}
	\label{check}
\begin{tabular}{c|c|c|c}
	\hline
	error bit & ${\bf c}$ & error bit & ${\bf c}$ \\
	\hline
	$b_1$ & 111 & $b_2$ & 110 \\
	$b_3$ & 101 & $b_4$ & 011 \\
	$b_5$ & 100 & $b_6$ & 010 \\
	$b_7$ & 001 & no & 000 \\
	\hline
\end{tabular}
\end{table}

Based on~\eqref{equ:check}, in this example, we calculate the syndrome as ${\bf c}= [1,1,1]$, which means the first bit is wrong. Therefore, we correct the information bitstream as ${\bm \hat{x}}= [0,0,1,0,1,0,1]$. The selected codeword index is $3$, which successfully correct the erroneous bit. In this way, by exploiting the ``self-correcting'' capabilities of channel coding, it is possible to obtain the correct angular index for beamforming even if the wide beam in the upper layer leads to incorrect decision. Thanks to the coding gain, it is promising that the proposed hierarchical beam training method can enable reliable beam training for remote users with low SNR.  

\subsection{Overall Idea Description} \label{Sec_4_Subsec_1}
In this subsection, we generalize the specific example of $(7,4)$ Hamming code to a unified coded beam training framework. 
We will first present the theoretical foundations of the proposed coded beam training and then the general coded beam training framework is illustrated.

\subsubsection{Theoretical Foundations}
The theoretical foundations of coded beam training lies in \emph{the duality of hierarchical beam training and channel coding}, which are elaborated below.


\begin{itemize}
\item \textbf{Channel coding}: A channel code consists of an encoder function $f$ and a decoder function $g$. The encoder $f$ maps a message $u\in \mathcal{U}$ to a codeword ${\bm x}=f(u)\in \mathcal{X}^n$, where $\mathcal{X}$ is the output alphabet of the encoder (usually binary, i.e., $\mathcal{X}=\{0, 1\}$), and $n$ is the code length. The channel $W: \mathcal{X}^n\to\mathcal{Y}^n$ randomly maps a coded sequence for transmission $\bm x$ to a received sequence $\bm y$, where $\mathcal{Y}$ is the received alphabet. Finally, the decoder $g$ maps $\bm y$ to an estimated message $\hat{u}\in \mathcal{U}$, which is expected to equal $u$ with high probability, i.e., $\mathbb{P}[u = \hat{u}]$ is close to $1$. The number of different messages $|\mathcal{U}|$ determines the number of payload bits $k=\log_2 (|\mathcal{U}|)$, and the code rate is defined as $R=k/n$. Thus, an $(n, k)$-code is a pair of encoder-decoder that takes $k$ bits into $n$ channel input symbols, and recovers $k$ bits from the $n$ output symbols of the channel. More details regarding the basics on channel codes can be found in~\cite{shannon1948mathematical}. 

\item \textbf{Beam training}: As mentioned in Section~\ref{Sec_3_Subsec_2}, the target of beam training is to determine the angular directions of the users from the received signals after the BS transmits a pre-designed beam training codebook $\mathcal{C}$. Generally, an $(M, N_{\rm T})$-{\it beam training code (BTC)} is defined as a beam training procedure capable of distinguishing $N_{\rm T}$ different angular directions via an $M$-layer beam training codebook. We denote the codebook  $\mathcal{C}=\{{{\mathcal{C}}(l,b)\in \mathbb{C}^{N_{\rm T}\times 1}, l\in \{1,2,\cdots,M\},b\in \{1,2,\cdots,b_l\}}\}$ with $b_l$ codewords for layer $l$. Besides, we denote the beam pattern corresponding to the codeword ${\mathcal{C}}(l,b)$ as $\mathcal{V}(l,b)\in \{0, 1\}^{N_{\rm T}}$, which describes the $0 \mbox{-} 1 $ pattern of the multi-mainlobe beam in the angular domain as is defined in Section~\ref{ham}. An information-theoretic insight is that it is only possible to construct $(M, N_{\rm T})$-BTC with $|M| = \Omega(\log N_{\rm T})$, since during beam training, the user can obtain only one bit of information during one training time slot. 
\end{itemize}

The above comparison reveals that beam training is intrinsically an information transmission problem. In the problem,  ``payload'' bits are the unknown physical direction of the UE, the ``channel'' is the angular response function of the physical channel, and the ``receiver'' is the UE itself. In this context, the BTC plays the same role as the channel codes during data transmission. 

From above description, we summarize the relationship between a BTC and a channel code: An $(n, k)$-channel code is equivalent to a $(n, 2^k)$-BTC. This relationship ensures that 
an arbitrary channel code can be converted to a reliable BTC to protect beam training in harsh channel conditions by introducing redundant bits, which motivates us to propose the following framework for channel code-BTC.

\subsubsection{Framework Description}
\begin{figure*}
	\centering 
	\includegraphics[width=\linewidth]{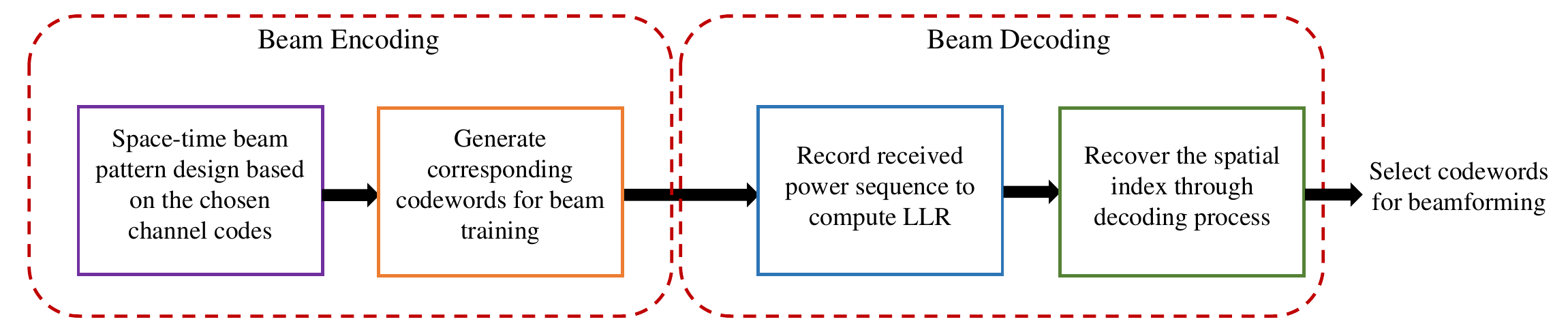}
	\caption{Overall framework of channel codes-aided hierarchical beam training with adaptive convolutional codes as an example. }
	\label{fig:framework}
\end{figure*}

As illustrated in Fig. \ref{fig:framework}, the framework of coded beam training comprises two stages. 
They are the \emph{beam encoding} for designing the BTC codebook, $\mathcal{C}$, and the \emph{beam decoding} to recover the users' spatial directions using $\mathcal{C}$, respectively. These two stages are detailed as follows. 

\emph{2.1) Beam encoding:} The target of beam encoding is to construct a $n$-layer BTC codebook, denoted as $\mathcal{C}$, from an $(n, k)$-channel code, which is capable of distinguishing $2^k$ angular directions. Similar to the design of $\mathcal{C}_{\rm ham}$ in the Hamming code case, each layer of the general codebook, $\mathcal{C}$, contains two complementary codewords, which are built on the following two steps: 
1) design the space-time beam pattern, $\mathcal{V}$, according to an arbitrary channel code, 2) generate the codebook $\mathcal{C}$ based on $\mathcal{V}$. 

\begin{itemize}

\item \textbf{Beam pattern design}: Recall that the set $\mathcal{V}$ records the ideal beam patterns of all codewords belonging to $\mathcal{C}$.
To achieve it, we index all the possible  angular directions $i\in [2^k]$ by a length-$k$ spatial information bits ${\bm u}_i \in \{0,1\}^{k}$, then the encoded bits ${\bm x}_i \in \{0,1\}^{n}$ is given by 
\begin{equation}
	{\bm x}_i = {f}({\bm u}_i) \in  \{0,1\}^{n}, \quad i\in [2^k],
\end{equation}
where $f(\cdot)$ denotes an arbitrary channel encoder. In this context, the space-time beam pattern $\mathcal{V}(l,b), l\in \{1,2,\cdots,n\},b\in\{1,2\}$ can be established according to the encoded bits as 
\begin{equation}
	\left\{\begin{array}{cc} \mathcal{V}(l,1,i) &= {\bm x}_i(l)\\
		\mathcal{V}(l,2,i) &= \bar{\bm x}_i(l) 
	\end{array} \right. , \quad l\in \{1,2,\cdots,n\}, \, i \in [2^k], 
\end{equation}
Consequently, the different spatial directions are encoded into different time sequences (i.e. different beam gain in the layer sequences), which is illustrated in Fig.~\ref{fig:scheme}. Note that $|\mathcal{C}|=2n$, i.e., the number of training time slots needed in the constructed BTC is $2n$. 

\item \textbf{Codebook generation}: Consider the generation of $\mathcal{C}$. 
Each codeword of $\mathcal{C}$, is optimized and generated by making its beam pattern as close to the ideal beam pattern, labeled by $\mathcal{V}$, as possible. This step can be efficiently performed using existing beam design methods~\cite{beamgene,10365224,7947209,10033092}, with the consideration of array structure constraints, such as the full-digital precoding and hybrid precoding. 
\end{itemize}


\emph{2.2) Beam decoding:} 
After acquiring the codebook $\mathcal{C}$,  the beam decoding can be performed to decode the spatial directions of the user with the received power sequence of the transmitted codewords in $\mathcal{C}$. 

 Specifically, 
 the BS sequentially assigns the beamformer $\bm w$ with ${\mathcal{C}}(l,1)$ and ${\mathcal{C}}(l,2)$ layer-by-layer and transmits pilots to the UE. Denote the UE's received signal power as $P(l,1)$ and $P(l,2)$. 
 To perform {\bf hard decoder}, UE compares the received power pair and records the results in a bit sequence $\hat{\bm x}$, i.e., $\hat{\bm x}(l) = 0$ if $P(l,1) > P(l,2)$ and  $\hat{\bm x}(l) = 1$ otherwise.  After the training phase, UE feeds back the bit sequence $\hat{\bm x}$ to the BS for carrying out hard decoding. Once receiving the feedback bit sequence $\hat{\bm x}$, BS can obtain $\hat{\bf u} = g(\hat{\bm x})$ as the estimation of the user's angular direction $\hat{i}$, by invoking the channel decoder $g(\cdot)$. 
In contrast,  if the {\bf soft decoder} is implemented, the UE needs to compute and record the log likelihood ratio (LLR) based on signal power sequence $P(l,1)$. And the estimation of the user's angular direction is calculated as  $\hat{\bf u} = g({\rm LLR})$.
 
In this way, the above constructed space-time beam pattern can distinguish $N_{\rm T} = 2^k$ different directions, which completes the beam decoding stage.



\ifx\onecol\undefined
\begin{figure}
	\centering 
	\includegraphics[width=\linewidth]{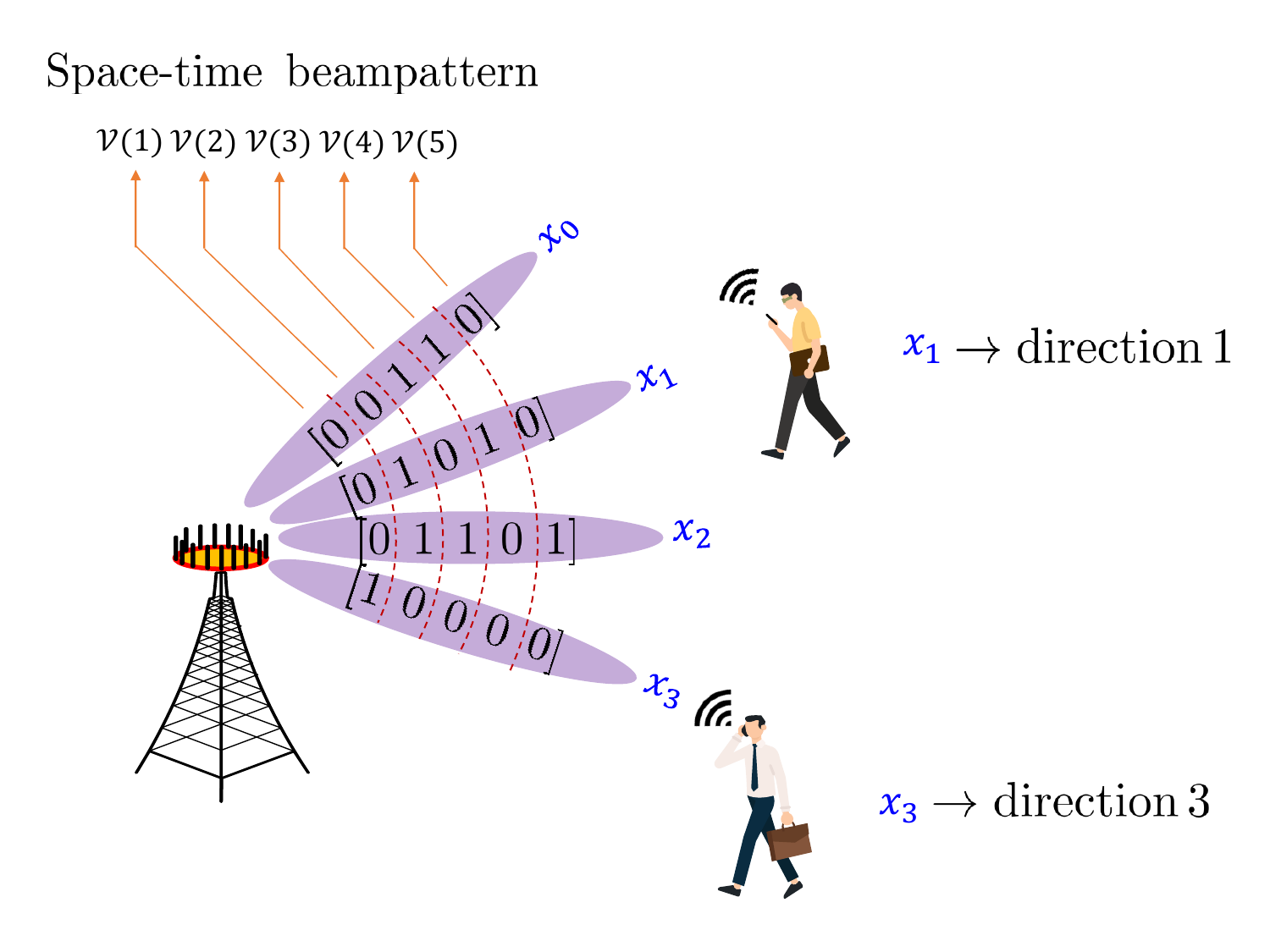}
	\caption{Space-time beam pattern: directions are encoded into $\bm x$ and then beam pattern $\mathcal{V}$ is constructed based on it.}
	\label{fig:scheme}
\end{figure}
\else 
\begin{figure}
	\centering 
	\includegraphics[width=0.6\linewidth]{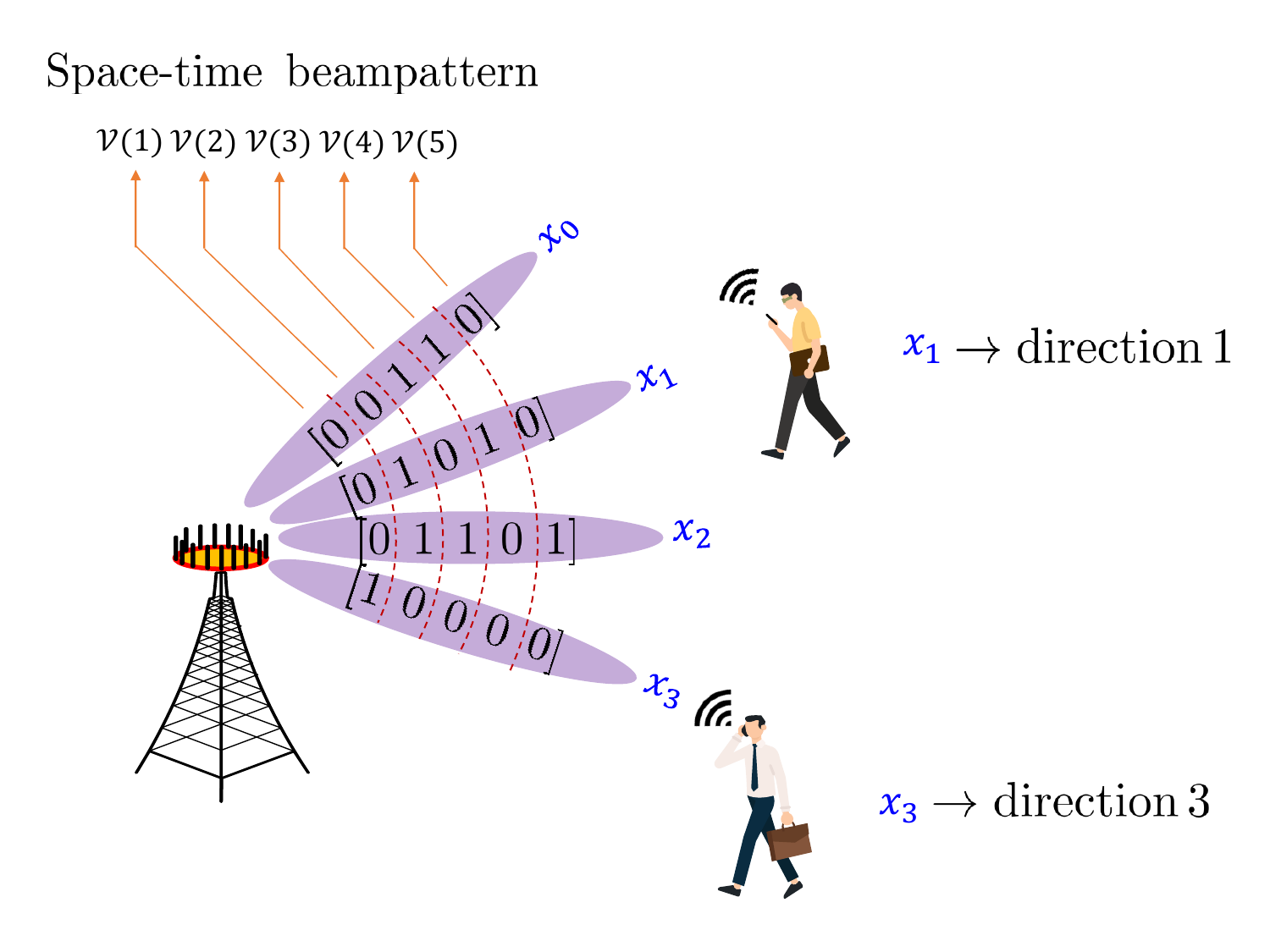}
	\caption{Space-time beam pattern.}
	\label{fig:scheme}
\end{figure}
\fi


\section{Implementation of Coded Beam Training} \label{Implementation}
The preceding example of $(7,4)$ Hamming code is only suitable to 16-antenna systems. To this end, this section considers the practical implementation of coded beam training. Specifically, we will first apply the convolutional channel code into the framework of coded beam training for communication systems with arbitrary number of antennas. Then, we also provide the extension of coded beam training to hybrid precoding architecture.

\subsection{Convolutional Beam Encoding} \label{Sec_4_Subsec_2}
Invented by P. Elias in 1955, convolutional codes are efficient error-correcting codes that have already been widely adopted in 4G LTE control channel coding standards~\cite{conv}. In the following sections, we employ convolutional codes as the channel coding method. 
\subsubsection{Convolutional Encoder}
The convolutional code is a coding scheme with memory that accepts a bitstream in blocks of length-$k$ and outputs a bitstream in blocks of length-$n$. Each block of $n$ output bits are determined by both the current input $k$ bits and preceding $N-1$ blocks, where $N$ represents the constraint length. Convolutional encoders are implemented by $N$ shift registers with taps determined by the generator polynomials. Here we adopt a convolutional encoder of $N=3, k=1, n=2$, as illustrated in Fig.~\ref{fig:encoder}~(a). Then, the output bits ${\bm x}(2i-1),{\bm x}(2i)$ can be computed as
\begin{equation}\label{eq:e1}
	{\bm x}(2i-1) = {\bf u}(i) \oplus {\bf u}(i-1) \oplus {\bf u}(i-2), 
\end{equation}
\begin{equation}\label{eq:e2}
	{\bm x}(2i)= {\bf u}(i) \oplus {\bf u}(i-2).
\end{equation}

\ifx\onecol\undefined
\begin{figure}
	\centering 
	\subfigure[Convolutional encoder for $N=3,k=1,n=2$. ]{
		\includegraphics[width=0.95\linewidth]{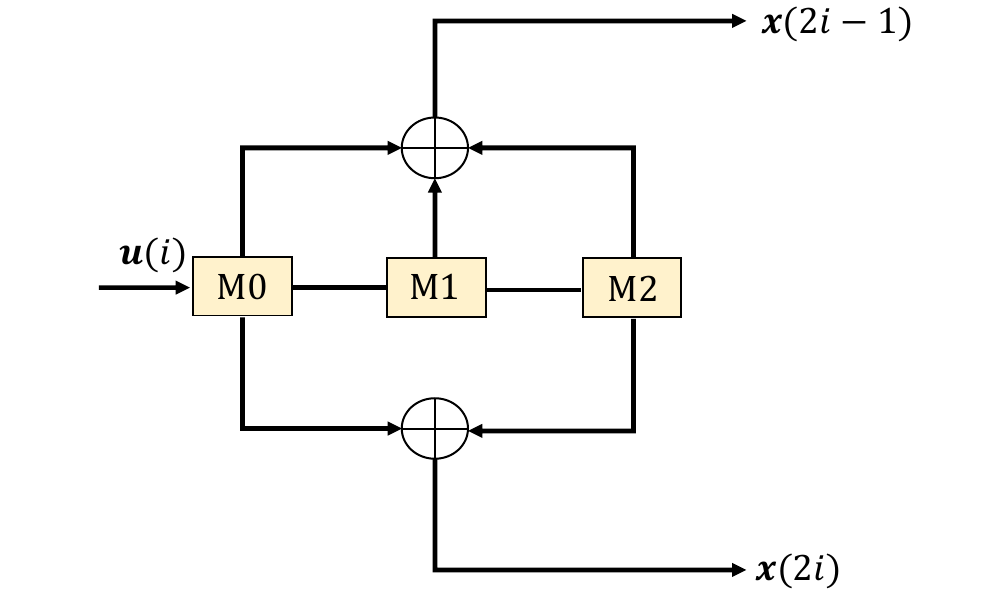}} 
	\subfigure[State transition diagram for convolutional encoder. ]{
		\includegraphics[width=0.95\linewidth]{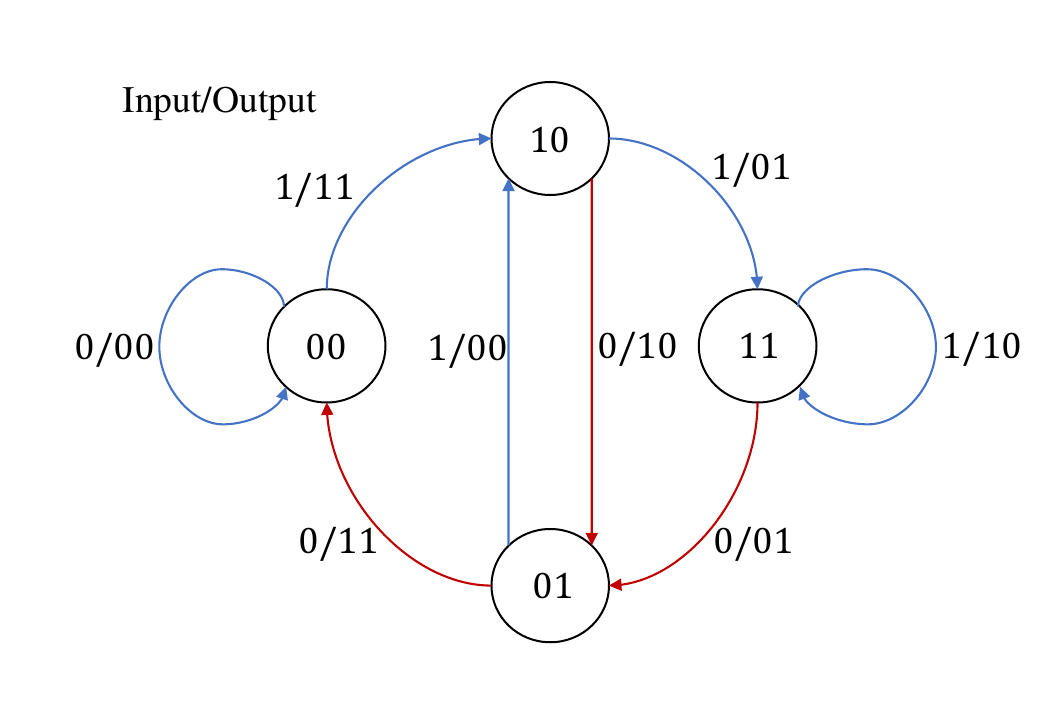}}
	\caption{Illustration of the convolutional encoder.}
	\label{fig:encoder}
\end{figure} 
\else 
\begin{figure}
	\centering 
	\subfigure[Convolutional encoder for $N=3,k=1,n=2$. ]{
		\includegraphics[width=0.5\linewidth]{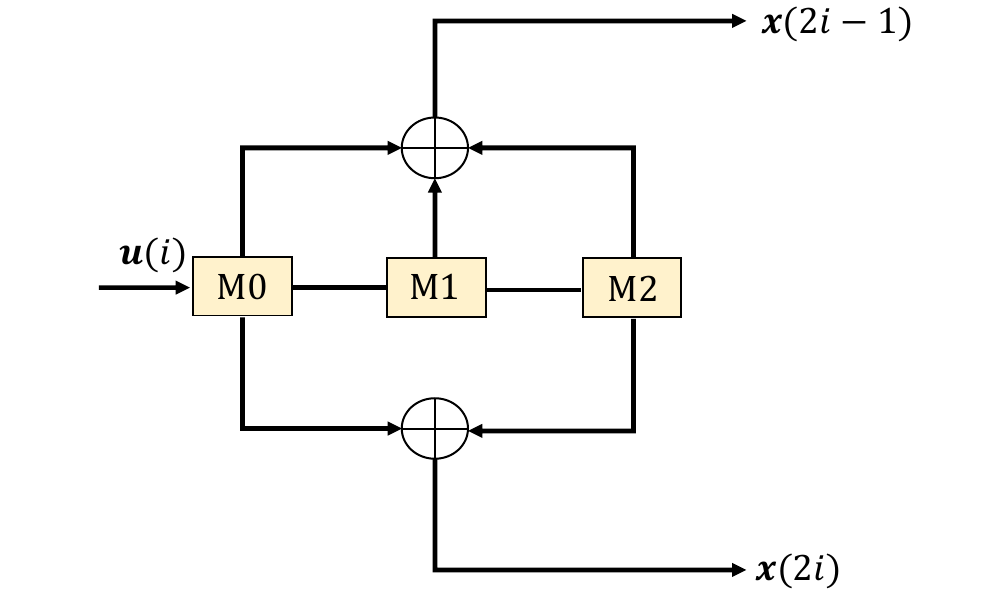}} 
	\subfigure[State transition diagram for convolutional encoder. ]{
		\includegraphics[width=0.5\linewidth]{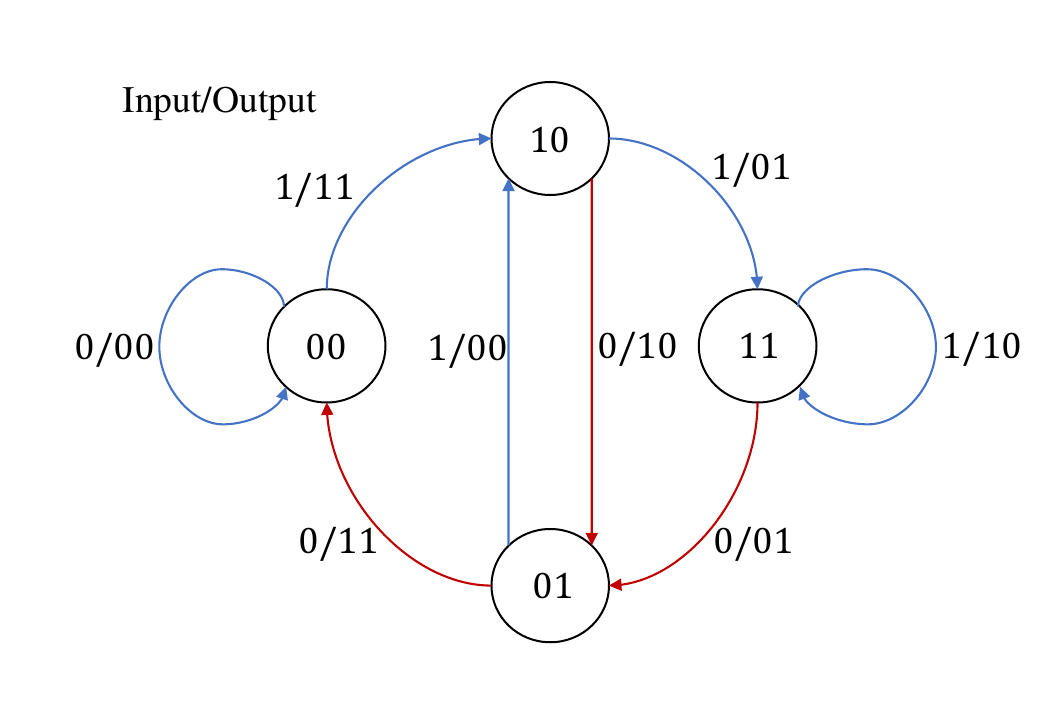}}
	\caption{Illustration of the convolutional encoder.}
	\label{fig:encoder}
\end{figure} 
\fi

The operation of the encoder proceeds as follows: Denote the bits ${\bf u}(i-1), {\bf u}(i-2)$ in the register $\rm M1, M2$ as ``state'' and initialize the state as $00$. Then the first input bit is fed into $\rm M0$ and outputs ${\bm x}(2i-1),{\bm x}(2i)$  according to~\eqref{eq:e1} and~\eqref{eq:e2}. Then, the next bit enters $\rm M0$ while the previous bits are shifted right for one bit. The process continues until eventually the last bit enters the register. The corresponding state transition diagram is presented in Fig.~\ref{fig:encoder}(b) and the entire process is denoted as $f_{\rm conv}$.

\subsubsection{Space-time Beam Pattern Design}
A key step of beam encoding is to design the space-time beam pattern in the BTC codebook. The complete hierarchical codebook $\mathcal{C}_{\rm conv}$ contains $M=2 \log_2 N_{\rm T}-1$ layers, consisting of  bottom layer and remaining upper layers. The remaining $M-1$ upper layers are designed based on the encoding algorithm of convolutional code $f_{\rm conv}$, each of which only includes one codeword since we utilize soft decoder instead of hard decoder to improve the performance. Besides, the bottom layer can be designed according to Eq.(\ref{bottom}), making use of high direction gain of codebook $\mathcal{W}$ to improve beam training performance. 
To design the space time beam pattern, we index all the possible angular directions with a bitstream of length $L=\log_2 N_{\rm T}-1$ and obtain the coded bit sequence as 
\begin{equation}
	{\bm x}_i = {f_{\rm conv}}({\bf u}_i) \in \{0,1\}^{M-1}, \quad i\in [2^L].
\end{equation}
Then, we construct the space-time beam pattern $\mathcal{V}_{\rm conv}(l), l\in \{1,2,\ldots ,M-1\}$ in codebook $\mathcal{C}_{\rm conv}$ according to the encoded bits as 
\begin{equation}
	\mathcal{V}_{\rm conv}(l,i) = {\bm x}_i(l), \quad i\in [2^L], 
\end{equation}

\subsubsection{Generate Corresponding Codewords}
Next, we focus on generating the codewords corresponding to the designed space-time beam pattern. In general, the normalized codewords of $l$-th layer $l \in \{1,2,\dots,M\}$ is denoted as $\mathcal{C}_{\rm conv} (l)$ and the corresponding complex beam gain vector of a multi-mainlobe beam $\mathcal{C}_{\rm conv} (l)$ with beam coverage $B_{l}$ is denoted as 
\begin{equation}
	{\bf g}_l = [g_l(\phi_1),g_l(\phi_1),\cdots,g_l(\phi_K)],
\end{equation}
where $K$ is the sampled angle number for beam generation. The beamforming gain can be presented as $g_l(\phi_n)=|g_l(\phi_n)|e^{\omega_n}$ where $\omega_n$ is phase information and the absolute beam gain $|g_l(\phi_n)|$ is predefined as 
\begin{equation}
	|g_{l}(\phi_n)|=\left\{\begin{array}{cc}
		\sqrt{2 / \vert B_{l} \vert}, & \phi_n \in B_{l} \\
		0, & \phi_n \notin B_{l}
		\end{array}\right.
\end{equation}
where $\vert B_{l} \vert$ is the coverage length of $B_{l}$~\cite{7845674}. In convolutional coding aided hierarchical codebook, the $B_{l}$ can be written as 
\begin{equation}
		B_{l}=\bigcup_{i} D_{i}, \text { if } \mathcal{V}_{\rm conv}(l,i)=0, i \in[2^{L}], \\
\end{equation}
where $D_{i}= [-1+i/2^{L-1},-1+(i+1)/2^{L-1}]\subset [-1, 1] $.

Obtaining the absolute beam gain vector we can generate the codewords based on GS codeword design algorithm proposed in ~\cite{10365224}. Employing the similarity of phase retrieval problem and codeword design \cite{gs}, GS-based codeword design procedure is shown in {\bf Algorithm ~\ref{alg:GS}}.
\begin{algorithm}[htb]
	\caption{ GS-based codeword design.}
	\label{alg:GS}
	\renewcommand{\algorithmicrequire}{\textbf{Input:}}
	\renewcommand{\algorithmicensure}{\textbf{Output:}}
	\begin{algorithmic}[1] 
	\REQUIRE $|{\bf g}|$, $I_{max}$, ${\bf A}$\\
	\STATE  Randomly initial phase information $w_n,n\in \{1,2,\cdots,K\}$ to obtain ${\bf g}^0$
	\FOR{each $i \in [1,I_{max}]$}
		\STATE   calculate ${\bf v}^i$ based on ${\bf g}^{i-1}$ according to (21)
		\STATE   ${\bf g}^{i}=|{\bf g}|\angle {\bf A}^H {\bf v}^i$
    \ENDFOR
	\STATE $\mathcal{C_{\rm conv}} = ({\bf A}{\bf A}^H)^{-1}{\bf A}{\bf g}^{I_{max}}$
	\ENSURE Designed codeword $\mathcal{C}_{\rm conv}$
	\end{algorithmic}
\end{algorithm}

Firstly, for each codeword, we randomly initial the phase information $w_n,n \in \{1,2,\cdots,K\}$ to obtain ${\bf g}^0$ (since the codeword generation is the same for each layer of codeword, we omit the subscripts of layer $l$ here). Then in the $i$-th iteration, ${\bf v}^i$ is calculated by least square algorithm as
\begin{equation}
	{\bf v}^i = ({\bf A}{\bf A}^H)^{-1}{\bf A}{\bf g}^{i-1}.
\end{equation}
In this way, the current complex beam gain can be written as $ {\bf A}^H {\bf v}^i$. In order to maintain the amplitude information of
desired $g$, we only utilize the phase information of current beam pattern $ {\bf A}^H {\bf v}^i$ to update ${\bf g}^{i}$. After
the iteration number reaches $I_{max}$, the designed codeword  $\mathcal{C}_{\rm conv}$ can be obtained. Futhermore, to fairly compare different codewords in each test, we usually normalize $\mathcal{C}_{\rm conv}(l)$ so that $\| \mathcal{C}_{\rm conv}(l) \|_2 =1$.

In the beam training process, the BS sequentially transmits $\mathcal{C}_{\rm conv}(l),l \in \{1,2,\ldots,M-1\}$ to the UE. Then the UE records the received signal power sequence $\mathcal{P} (l)$ for beam decoding in the following section.


\subsection{Convolutional Beam Decoding} \label{Sec_4_Subsec_3}
The objective of beam decoding is to select the optimal codeword in $\mathcal{W}$. Based on the received signal power sequence, we are capable of determining whether the UE is in the coverage $B_{l}$ of $\mathcal{C}_{\rm conv} (l)$, and thus recover the spatial information to select codeword with the aid of convolutional decoding algorithm. 

\subsubsection{Viterbi Decoder}
There are a variety of algorithms for decoding the received power sequence, among which Viterbi algorithm is an effective and practical technique~\cite{DAVIDFORNEY1974222}. Exploiting dynamic programming~\cite{DAVIDFORNEY1974222}, the Viterbi decoder works in a sequential manner, where the output $\rm LLRs$ of the noisy channel are fed into a trellis graph. Then, sequential maximum a posteriori (MAP) estimators are applied to this trellis graph, in order to retrieve the most possible information sequence. Fig.~\ref{fig:trellis} illustrates the trellis of the utilized decoder with the initial state being $00$.

\ifx\onecol\undefined
\begin{figure}
	\centering 
	\includegraphics[width=\linewidth]{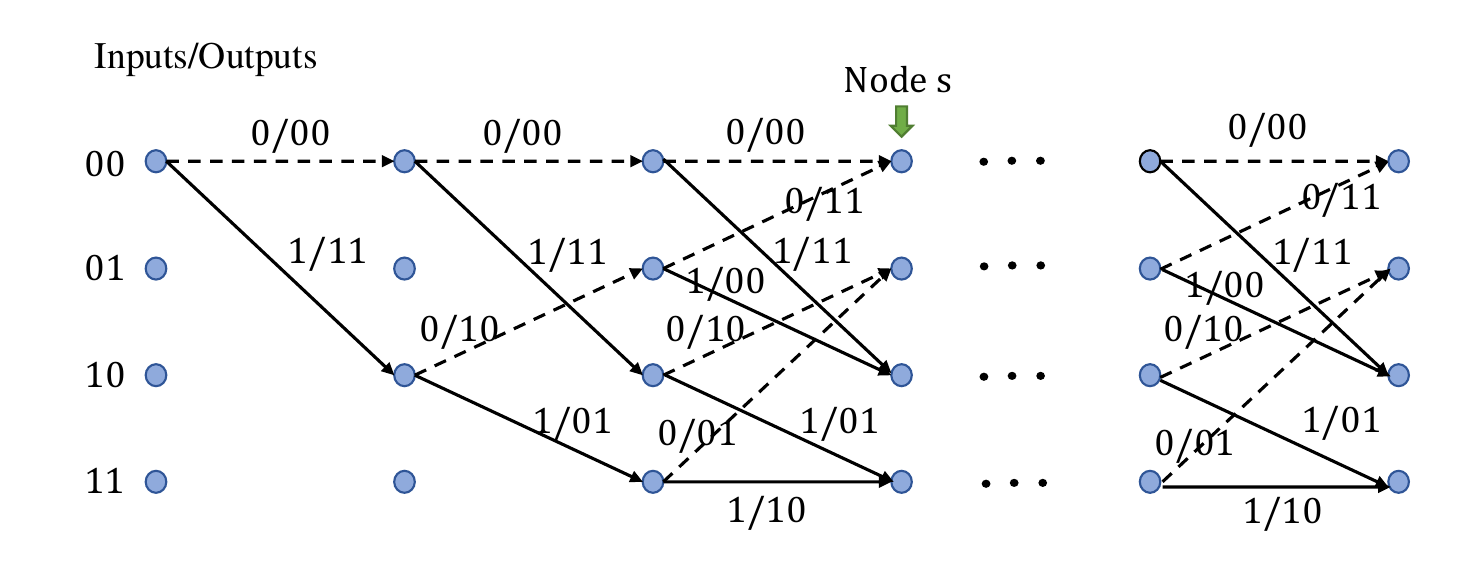}
	\caption{Trellis of the convolutional decoder. }
	\label{fig:trellis}
\end{figure}
\else 
\begin{figure}
	\centering 
	\includegraphics[width=0.8\linewidth]{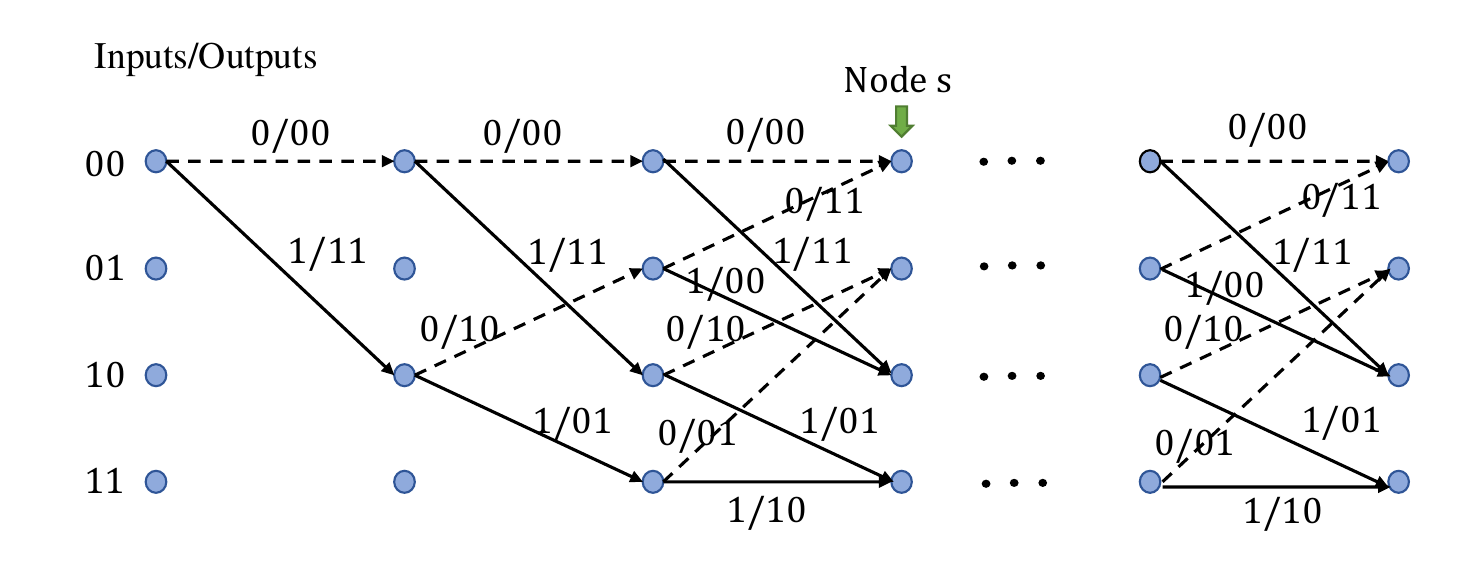}
	\caption{Trellis of the convolutional decoder. }
	\label{fig:trellis}
\end{figure}
\fi

\paragraph{Calculate $LLR$}
As discussed above, the critical component to the efficiency of convolutional decoder is the accurate computation of ${\rm LLR}$. Therefore, in this paragraph, we will focus on the calculation of ${\rm LLR}$.

If the UE is in the coverage $B_{l}$ of $\mathcal{C}_{\rm conv} (l)$, the ideal power of received signal is $\mathcal{P}(l)=\vert A_l+n \vert^2$. The $A_l$ is the ideal received signal power determined by the coverage length of $B_{l}$. In contrast, if the UE is \emph{not} in the coverage $B_{l}$ of $\mathcal{C}_{\rm conv} (l)$, the received signal power is  $\vert n \vert^2$. It is worth noting that traditional decoding algorithms are designed based on Gaussian channels, while the probability distribution of $\mathcal{P}$ obey $\mathcal{X}^2$ distribution rather than Gaussian distribution. Therefore, existing decoding algorithm can not be directly employed. The conditional probability density function of $\mathcal{P}(l)$ can represented as 
\begin{equation}
	p(\mathcal{P}(l)=x \vert \theta_{\rm UE} \in B_{l})= \frac{1}{\sigma^2} e^{-\frac{x+A_l^2}{\sigma^2}} {I_0} \left (\frac{\sqrt{A_l^2 x}}{\sigma^2/2} \right )
\end{equation}
\begin{equation}
	p(\mathcal{P}(l)=x \vert \theta_{\rm UE} \notin B_{l})= \frac{1}{\sigma^2} e^{-\frac{x}{\sigma^2}} 
\end{equation}
where ${I_0}$ is zeroth order modified Bessel function of the first kind, $A$ is the received amplitude. Therefore, the ${\rm LLR}$  can be expressed as 
\begin{eqnarray}
	{\rm LLR}  &=& \log \frac{p( \mathcal{P}(l)=x \vert \theta _{\rm UE} \in B_{l})}{p( \mathcal{P}(l)=x \vert \theta _{\rm UE} \notin B_{l} )}\\
	    &=& -\frac{A_l^2}{\sigma^2} + \log {I_0}\left (\frac{\sqrt{A_l^2 x}}{\sigma^2/2} \right) \nonumber
\end{eqnarray}\label{eq:llr}

After obtaining ${\rm LLR}$, beam decoding can be performed to recover the information bits through the Viterbi decoder, which is specified in next paragraph. 

\paragraph{Recover the Spatial Index Through Decoding Process}
The Viterbi algorithm-based beam decoding process proceeds in a step-by-step fashion as follows:

For \textbf{initialization} step, set initial loss of survivor paths ${\rm loss}_{0} \in \mathbb{R}^{1\times 4} $ as $\bf 0$ and all survivor paths as empty. In the \textbf{training} process, for $n=2$ received power $P(2l-1),P(2l),l\in \{1,2,\ldots L\}$ in the $l$-th level, compute the ${\rm LLR}$ as $llr1_l$ and $llr2_l$ according to Eq.(25). 

Denote the two coded sequences (i.e. outputs of channel coding) of the paths entering the node $s$ as $\bm{y_{s1}}$ and $\bm{y_{s2}}$ and the corresponding incoming nodes as node $t_1$ and $t_2$, respectively. For example, for the node $s$ in Fig.~\ref{fig:trellis}, the incoming nodes are $t_1=1$ and  $t_2=2$ while the coded sequences of entering paths are   $\bm{y_{s1}}=00$ and $\bm{y_{s2}}=11$. Then UE can compute the ``distance'' for two paths entering each state of the trellis by adding the ``distance'' of incoming branches to the ``distance'' of the connecting survivor path from incoming node $t$ level $l-1$ as 
\begin{equation}
	l_1=  {\rm loss}_{l-1}(t_1) + (-1)^{\mathbb{I} ({\bf y_{s1}}(1)=0)}llr1 + (-1)^{\mathbb{I} ({\bf y_{s1}}(2)=0)}llr2
\end{equation}
\begin{equation}
	l_2=  {\rm loss}_{l-1}(t_2) + (-1)^{\mathbb{I} ({\bf y_{s2}}(1)=0)}llr1 + (-1)^{\mathbb{I} ({\bf y_{s2}}(2)=0)}llr2
\end{equation}
where $\mathbb{I}(\cdot)$ in the indicator function. According to  maximum a posteriori estimators, UE can select the lowest ``distance'' as  the ${\rm loss}_{l}(s)$ of survivor path for node $s$ in level $l$, which can be presented as
\begin{equation}
	{\rm loss}_{l}(s) = \min \{l_1,l_2\},
\end{equation}
and  the selected node is node $t^*$.
Denote the input bit corresponding to the selected incoming path as $b(s) \in \{0,1\}$, which will be feedback to the BS. Then the BS updates the survivor paths as 
\begin{equation}
	{\rm path}_l(s)={\rm append}({\rm path}_{l-1}(t^*),b(s))
\end{equation}

Continue the computation until the algorithm completes its forward search. Then the BS can select the node with lowest ``distance'' at the terminal level and the corresponding survivor path as ${\bf q}$. Through this way, BS can obtain a decimal index $\mathcal{T}={\rm bintodec}({\bf q})$ which includes two codewords in the bottom layer.  Lastly, in the bottom layer, we test two codewords of index $2\mathcal{T}+1$ and $2\mathcal{T}+2$ in codebook $\mathcal{W}$ to acquire the final selected codeword.
\paragraph{ML Decoder}
Different decoding algorithms can result in different beam training performances, therefore to intuitively evaluate the effectiveness of our improved decoder we attempt to derive the performance bound of convolutional code. Maximum likelihood (ML) decoding is the optimal decoding method that minimizes the probability of decoding errors when each codeword is sent with an equal probability. The computational complexity of ML decoder prohibits it from practical employment since the required computation complexity grows exponentially with the input length. However, it serves as the performance bound of convolutional codes. For the first $M-1$ layers, ML decoder selects the UE direction index $idx=i$ with the maximal probability of received signal ${\bm{x}}$, i.e.
\begin{equation}
	i = \max \limits_{i} p({\bm{x}}|i)
\end{equation}
Let $\mathcal{N}_{0i} = \{l|\mathcal{V} (l,i)=0,l\in \{1,2,\ldots,M-1\}\}$ be the set where the beam pattern is $0$, while $\mathcal{N}_{1i} = \{l|\mathcal{V} (l,i)=1,l \in \{1,2,\ldots,M-1\}\}$ be the set where the beam pattern is $1$. Therefore, $p({\bm{x}}|i)$ can be expressed as
\begin{equation}
	\vspace{0.01in}
\begin{split}
	p({\bm{x}}|i)  &= \prod \limits_{l \in \mathcal{N}_{0i}} \frac{1}{\sigma^2} e^{-\frac{x_l}{\sigma^2}} \\ &\cdot  \prod \limits_{l \in \mathcal{N}_{1i}} \frac{1}{\sigma^2} e^{-\frac{x_l+A_l^2}{\sigma^2}} {I_0}\left(\frac{\sqrt{A_l^2 x_l}}{\sigma^2/2}\right) 
\end{split}
\end{equation}
Thus, the log likelihood is
\begin{equation}
	\begin{split}
	\log   &  p({\bm{x}}|i)= -2N \log \sigma - \sum \limits_{l\in \mathcal{N}_{0i}}   \frac{x_l}{\sigma^2}- \sum \limits_{l\in \mathcal{N}_{1i}}   \frac{x_l+A_l^2}{\sigma^2} \\&+ \sum \limits_{l\in \mathcal{N}_{1i}} \log {I_0}\left(\frac{\sqrt{A_l^2 x_l}}{\sigma^2/2}\right) \\ &= -2N \log \sigma - \frac{N_{i1} A_l^2+ \sum x_l}{\sigma^2} + \sum \limits_{l\in \mathcal{N}_{1i}} \log {I_0}\left(\frac{\sqrt{A_l^2 x_l}}{\sigma^2/2}\right) 
	\end{split}
\end{equation}
where $ N_{1i}=|\mathcal{N}_{1i}|$ is the number of elements in the set $\mathcal{N}_{1i}$. Then the ML decoder then can be simplified as
\begin{equation}\label{equ_ml}
	i = \max \limits_{i} \sum \limits_{l\in \mathcal{N}_{1i}}  -\frac{A_l^2}{\sigma^2}+ \log {I_0}\left(\frac{\sqrt{A_l^2 x_l}}{\sigma^2/2}\right) 
\end{equation}
Based on (\ref{equ_ml}), we sequentially test all $2^L$ direction indexes to select the optimal index $i$ and then acquire the performance bound of convolutional decoder. The more the performance of a designed decoding algorithm approaches that of ML decoder, the more efficient the designed decoder is.


\subsection{Adaptive Beam Encoding based on Decoding Algorithm}
The codebook proposed in Section.~\ref{Sec_4_Subsec_2} consist of multi-mainlobe beams with  the coverage length of $\vert B_{l} \vert$. Though we error correction capabilities can help resolve the  ``error propagation'' dilemma, the reliability of beam training performance for UEs may be limited by the low directional gain of wide beams. Fortunately, according to Section.~\ref{Sec_4_Subsec_3}, the Viterbi decoder gradually ``truncates'' impossible paths and only retains survivor paths. Therefore, in this section, we propose an adaptive beam encoding algorithm which gradually increases the beam gain with the aid of  Viterbi decoder. In this context, energy is focused only within the directions corresponding to the survivor paths.

First we design the space-time beam pattern according to Eq.(15)-(16), which is the same as Section.~\ref{Sec_4_Subsec_2}. 	For the designed beam in the $l$-th layer, we only require to inject energy to the the directions corresponding to the survivor paths in layer $l-1$. Transfer the bitstream ${\rm path}_{l-1}(s),s\in \{1,2,3,4\}$ to decimal index $d_{l-1}(s)$, and the indicated direction corresponding to the path is
\begin{equation}
	{\rm Dir}_{l-1}(s) = [-1+d_{l-1}(s)/2^{l-2},-1+(d_{l-1}(s)+1)/2^{l-2}].
\end{equation}
Then the survivor direction for layer $l$ can be expressed as
\begin{equation}
	S_l = \bigcup_{s} {\rm Dir}_{l-1}(s), s\in \{1,2,3,4\},
\end{equation}
and based on it we can adjust the coverage of beam in layer $l$ as
\begin{equation}
	{\rm Bnew}_l = B_l \bigcap S_l.
\end{equation}
Then we generate the corresponding codewords based on GS algorithm. We summarize the adaptive coded beam training framework as Fig.~\ref{fig:process}.

\ifx\onecol\undefined
\begin{figure}
	\centering 
	\includegraphics[width=\linewidth]{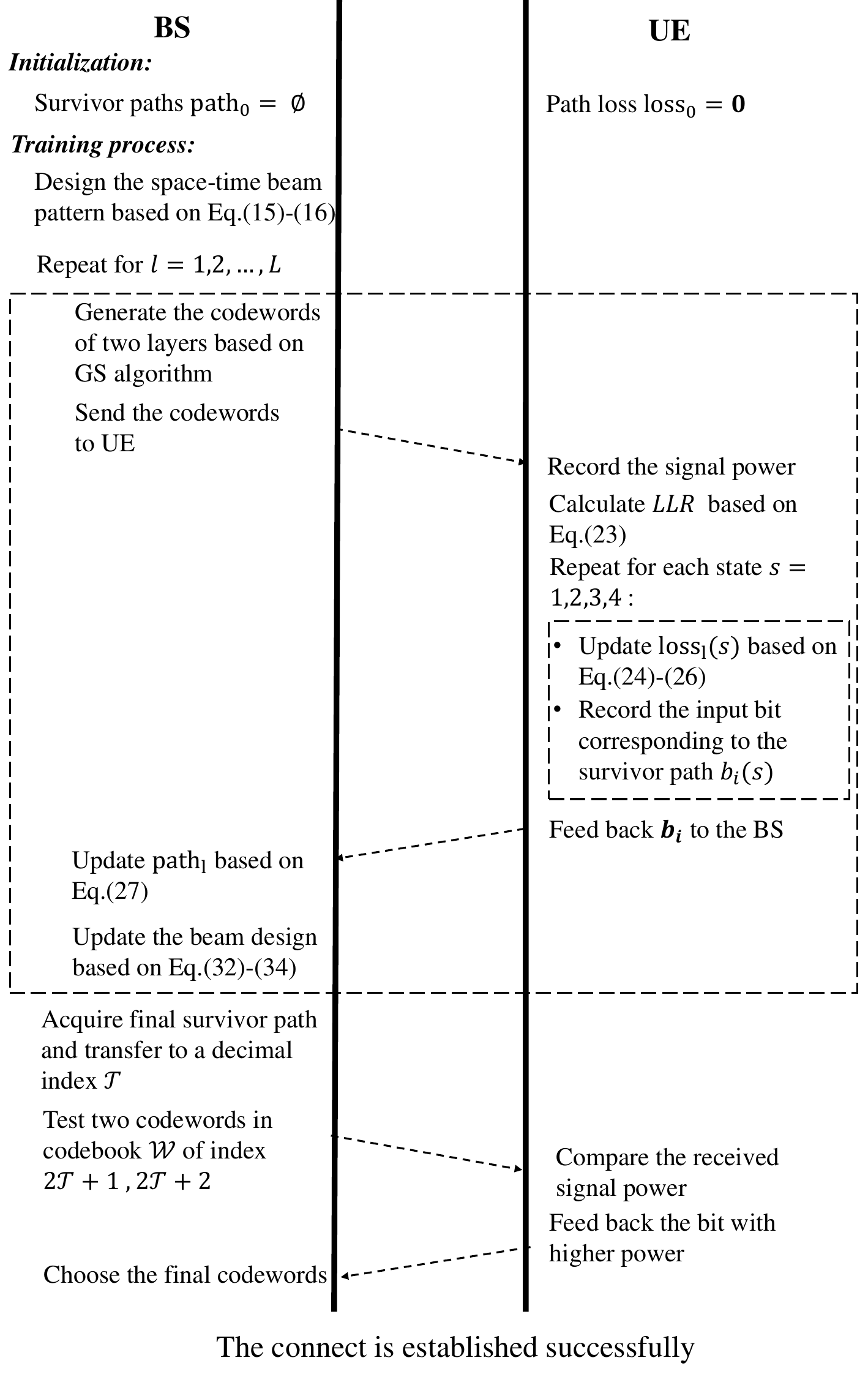}
	\caption{Signaling diagram of the coded beam training procedure. }
	\label{fig:process}
\end{figure}
\else 
\begin{figure}
	\centering 
	\includegraphics[width=0.6\linewidth]{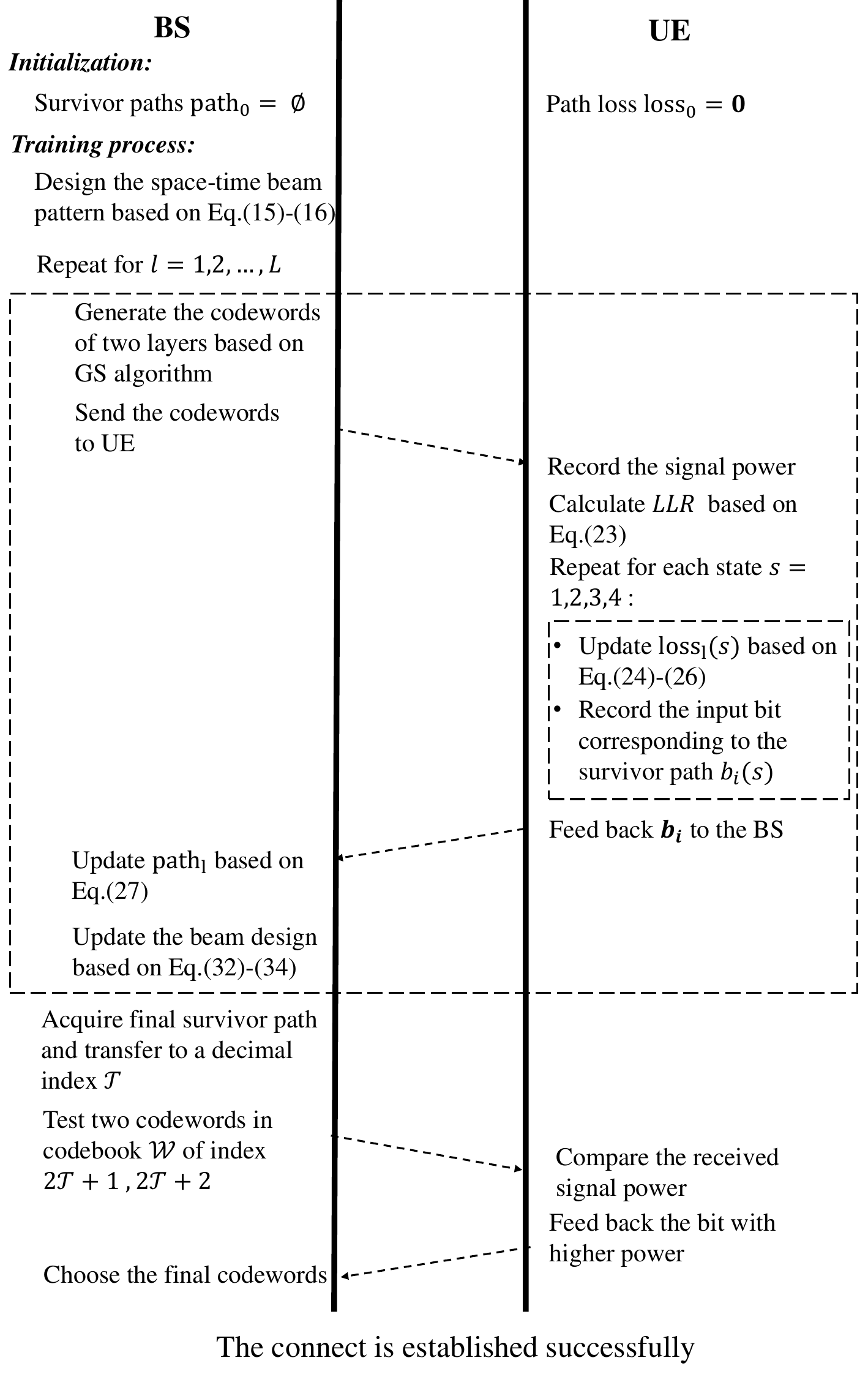}
	\caption{Signaling diagram of the coded beam training procedure. }
	\label{fig:process}
\end{figure}
\fi

\subsection{Extension to Hybrid Precoding Structure}\label{Sec_4_Subsec_4}
As we have clarified in Section.~\ref{Sec_3_Subsec_1}, the proposed method is independent of the precoding architecture, so we utilize full-digital precoding scheme for concise  representation. In this subsection, we will demonstrate how the proposed method can be conveniently transfered to hybrid precoding scheme, which is widely employed in exsiting communication systems.

Consider a typical mmWave/Terahertz massive MIMO system where the BS employs $N_{\rm RF}(N_{\rm RF}\ll N_{\rm T})$ RF chains to serve a single-antenna user. The BS employs hybrid precoding, and the  optimization problem can be decomposed into two sub-problem: digital precoding optimization and analog precoding optimization. For analog precoding, the training process is the same with that of full-digital structure. The codewords chosen finally in coded beam training meet the requirements of constant envelop constraint due to phase shifters. The only difference lies in that the codewords required during beam training should be generated in hybrid structure insetead of full-digital structure. As for this issue, the authors in~\cite{Hybrid} have verified that the full-digital structure can be approximate by hybrid precoding with $N_{\rm RF}>2N_s$ RF chain where $N_s$ denotes the data stream number. Besides, several beamformers~\cite{tdma,8408778,9411813} have been proposed to generate required wide beams with hybrid structure. Therefore, the proposed method can be directly transfered to analog beamformer design.

After obtaining the analog beamformer, we design the digital precoding based on the low complexity Zero Forcing (ZF) algorithm~\cite{marzetta2010noncooperative}.

\subsection{Extension to Multi-User Scenarios}
The above proposed model supposes single-user communication scenarios for clear expression, and in this subsection we aim to reveal the scalability of proposed coded beam training framework to multi-user scenarios. 

The space-time beam pattern and codeword generation process described in Section~\ref{Sec_4_Subsec_2} is independent of UE, so the BS can send the same codewords to different users. Then each user performs Viterbi decoding to find the survivor paths simultaneously by their own. As for the adaptive beam encoding where we adapt the beam design according to the feed back during the training process, the only difference is that we are supposed to inject energy to angular directions of the \textbf{union} of survivor paths for each user.

\section{Simulation Results} \label{sec-re}

\subsection{Complexity Analysis} \label{Sec_4_Subsec_5}
In the proposed method, the BS transmits a single codeword to the UE for upper $M-1$ layers of $\mathcal{C}$, which occupies $M-1$ time slots. At the bottom layer of $\mathcal{C}$, the BS sends two codewords to the UE, culminating in the determination of the ultimate chosen codewords, which takes up $2$ time slots. Therefore, the proposed beam training scheme takes up $M+1=2\times \log_2 N_{\rm T}$ time slots. We summarize the beam training overheads of our proposed coded beam training method, exhaustive beam sweeping method and binary search-based hierarchical beam training~\cite{tdma}, as shown in Table.~\ref{tab:com}. Suppose $N_{\rm T}=1024, M=19$, the training overheads of our proposed method, exhaustive beam sweeping method and binary search-based hierarchical beam training are $20, 1024$ and $20$ time slots, respectively. Our proposed method exhibits training overheads comparable to  binary search-based hierarchical beam training and substantially curtails training  overheads by $98\%$ compared with exhaustive beam sweeping.

We also conduct a comparison of the feedback overhead from the UEs to the BS. In the proposed fixed coded beam training method in Section.~\ref{Sec_4_Subsec_2} and Section.~\ref{Sec_4_Subsec_3}, the method requires one time slot to feedback the decoded angular direction after codewords in $M-1$ layers are all transmitted. Then after the beam training for the bottom layer, an additional time slot is expended to feed the beam index back to BS. Therefore, the number of the cumulative feedback time slot is $2$. In contrast,  for the adaptive coded beam training method, BS  necessitates  the feedback from the UE to dynamically adjust the beam pattern every two layers. In such cases, the time slots needed amount to  $\log_2 N_{\rm T}-1$. Adding the time slot required in the bottom layer, there are $\log_2 N_{\rm T}$ time slots required in total. It  is consistent with  binary search-based hierarchical beam training which also relies on  the feedback from UE to choose codewords for the subsequent layer within the codebook of length $\log_2 N_{\rm T}$. For exhaustive beam sweeping, UE the subsequent layer within, which results in totally $1$ times of feedback.

\begin{table*}[htb]
	\centering
	\caption{Comparisons of overheads for different Schemes}
	\begin{tabular}{ccc}
		\toprule
		Schemes & Training Overheads & Feedback Overheads \\
		\midrule
		Adaptive coded beam training & $2 \log_2 N_{\rm T}$ & $\log_2 N_{\rm T}$ \\
		Fixed coded beam training & $2 \log_2 N_{\rm T}$ & $2$ \\
		Exhaustive beam sweeping & $N_{\rm T}$ & $1$ \\
		Binary search-based hierarchical beam training & $2\log_2 N_{\rm T}$ & $\log_2 N_{\rm T}$ \\
		\bottomrule
	\end{tabular}
	\vspace{0.5em}
	\label{tab:com}
	\vspace{-2em}
\end{table*}

\subsection{Performance Analysis}
In this section, numerical results are presented to evaluate the performance of the designed coded beam training framework, and we draw a comparison with both exhaustive beam sweeeping and exsiting hierarchical beam training methods. The simulations are mainly based on the adaptive coded beam trianing.

We consider an XL-MIMO communication system, where a single-antenna user is served by a BS. The BS is equipped $1024$-antenna ULA, with spacing between antennas equal to $\lambda/2$ in digital precoding system. We adopt the Saleh-Valenzuela channel model described in Eq.~\eqref{eq-channel} with LoS component being the dominant path. We further assume the UEs are uniformly distributed within physical direction $[0,2\pi]$. The performances of success rate and achievable rate are all averaged on the instantaneous results of 1000 random Monte Carlo realizations of channel. 

\ifx\onecol\undefined
\begin{figure}
	\centering 
	\includegraphics[width=\linewidth]{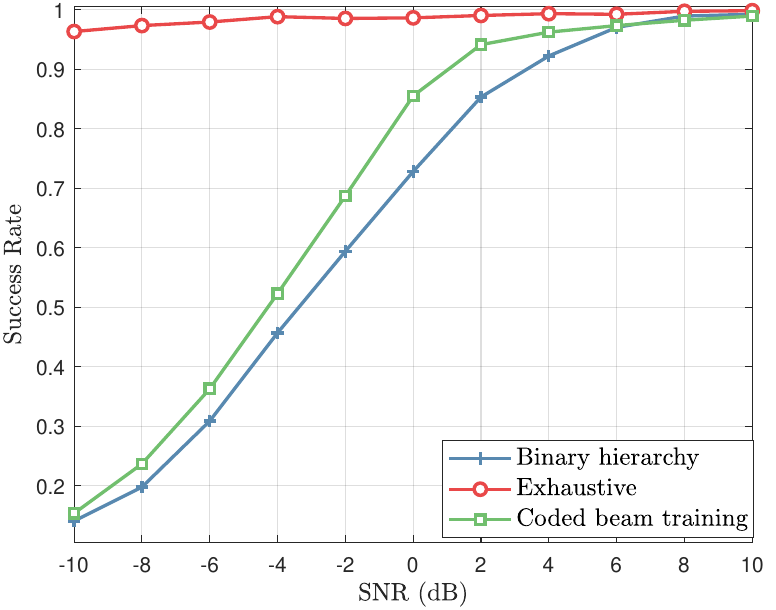}
	\caption{Comparisons of the success rate for different beam training methods.}
	\label{fig:successrate}
\end{figure}
\else 
\begin{figure}
	\centering 
	\includegraphics[width=0.6\linewidth]{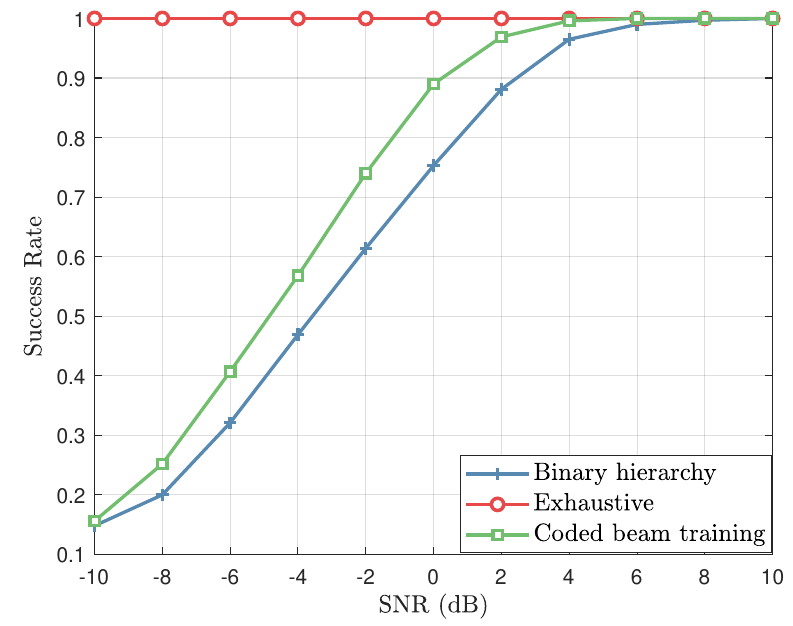}
	\caption{Comparisons of the success rate for different beam training methods.}
	\label{fig:successrate}
\end{figure}
\fi

Fig.~\ref{fig:successrate} depicts a comparison of the success rate of beam training for different schemes. For fair evaluation, we emphasize that the proposed coded beam training method shares equivalent training overheads with that of binary search-based hierarchical beam training. In such a case, the performance gain can be attributed to the coding gain  facilitated by channel codes, rather than  an increased utilization of pilot resources. It is evident that the scheme in~\cite{exhaustive} attains a superior performance than the other schemes, which lies in the fact that the exhaustive beam sweeping, whose training overhead is much higher than the other schemes, inherently perform better at the expense of training efficiency. Existing hierarchical beam training method significantly reduces the training overheads, but it is not capcable of obtaining reliable beam training performance for remote users with low SNR. This inadequacy arises from the ``error propagation'' phenomenon with the lower signal power of wide beam during beam training. Our proposed method significantly improves the success rate compared to existing beam training method, especially at low SNR while maintaining training overhead low, thanks to the leverage of the error correction capability of channel codes.

\ifx\onecol\undefined
\begin{figure}
	\centering 
	\includegraphics[width=\linewidth]{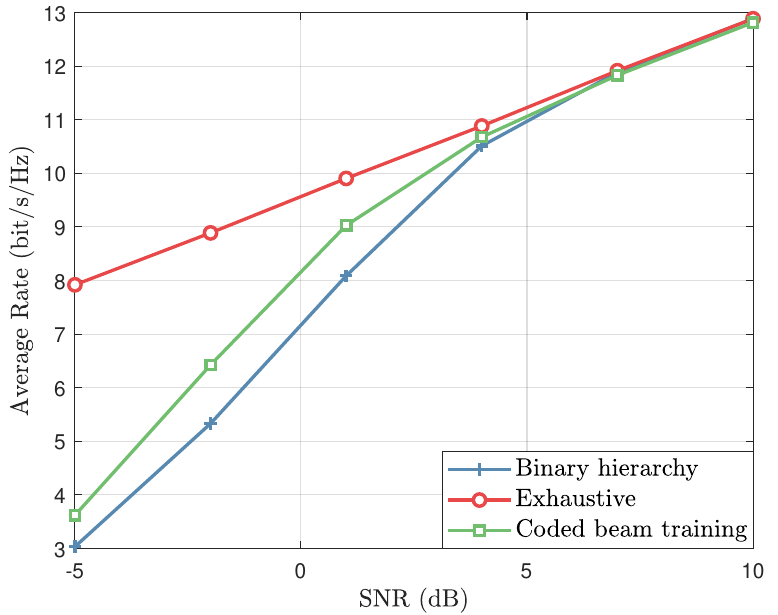}
	\caption{Comparisons of the average rate for different beam training methods. }
	\label{fig:sumrate}
\end{figure}
\else 
\begin{figure}
	\centering 
	\includegraphics[width=0.6\linewidth]{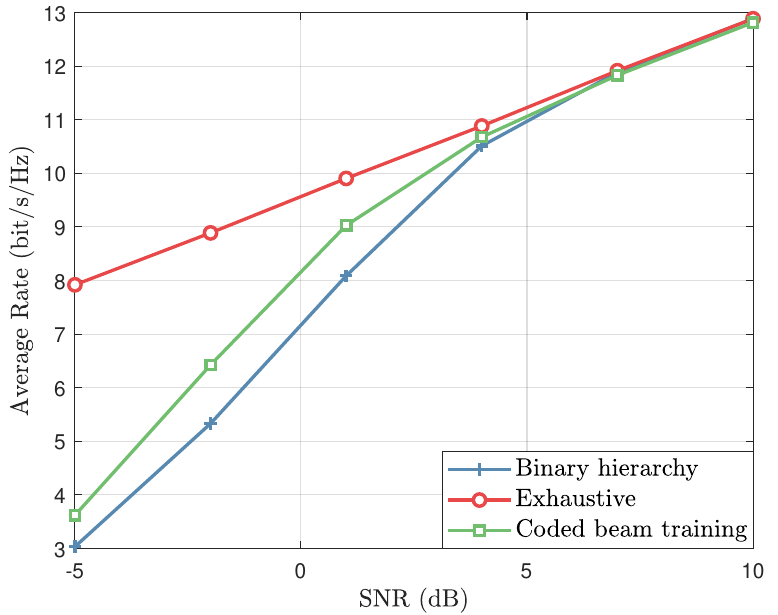}
	\caption{Comparisons of the average rate for different beam training methods. }
	\label{fig:sumrate}
\end{figure}
\fi

Fig.~\ref{fig:sumrate}  offers a comparative view of the achievable rate for different beam training schemes. The graph clearly illustrates the performance of our proposed method outstands traditional hierarchical beam training method with comparative training overheads, especially at low SNR. Moreover, as the SNR increases, the performance gap between our scheme and the exhaustive beam sweeping scheme in~\cite{exhaustive} diminishes, where the curves of our scheme and the beam sweeping scheme almost coincide at ${\rm SNR = 6\, dB}$. However, it's worth noting that the proposed method achieves considerable reduction (more than 98\% reduction) in training overhead. Therefore, we can conclude that our scheme strikes a remarkable balance between training overhead and beam training performance.

\ifx\onecol\undefined
\begin{figure}
	\centering 
	\includegraphics[width=\linewidth]{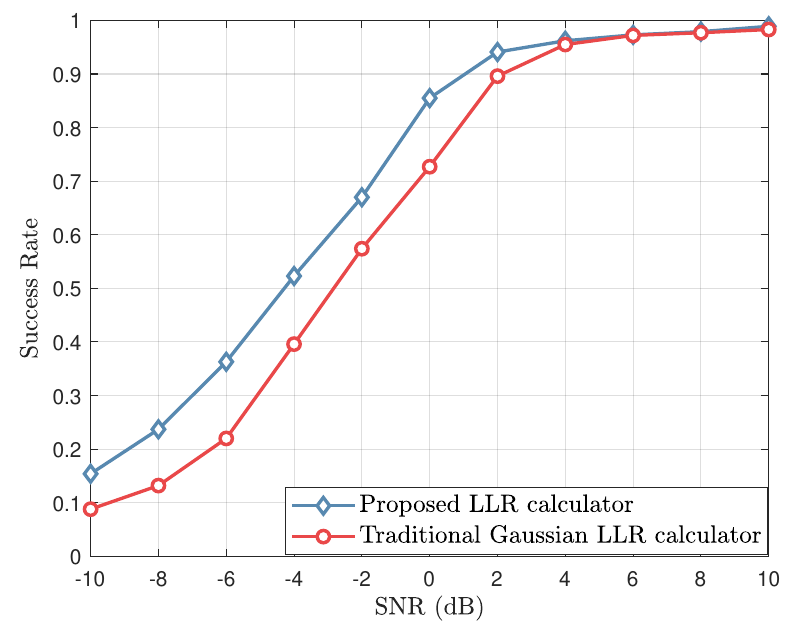}
	\caption{Comparisons of the performance for convolutional decoders with different LLR calculators.}
	\label{fig:ml}
\end{figure}
\else 
\begin{figure}
	\centering 
	\includegraphics[width=0.6\linewidth]{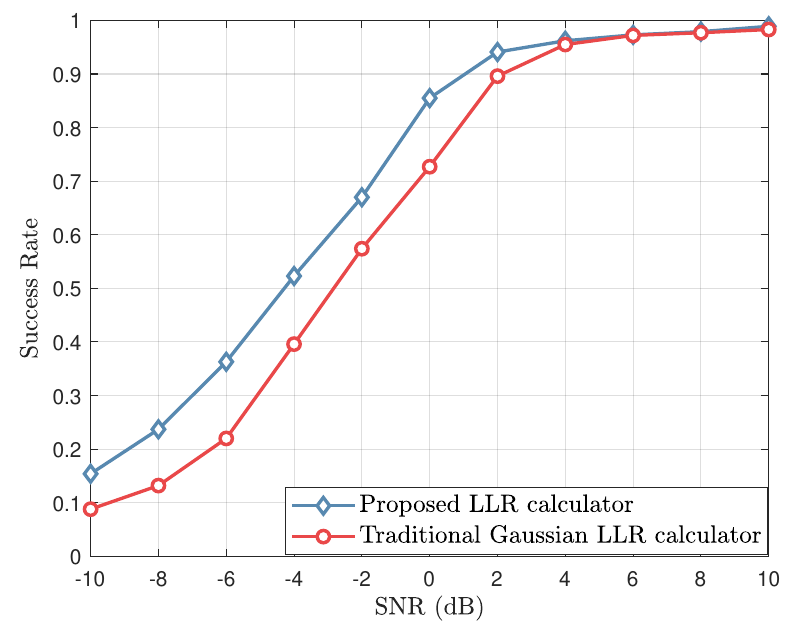}
	\caption{Comparisons of the performance for convolutional decoders with different LLR calculators.}
	\label{fig:ml}
\end{figure}
\fi

Furthermore, we verify the effectiveness of proposed enhanced convolutional decoding algorithm. Fig.~\ref{fig:ml} reveals the performance of proposed method and scheme with traditional decoder. The simulation results distinctly highlight the superior performance of our improved decoder in contrast to the traditional Gaussian distribution-based decoder, which substantiates the practicability of our modified decoding algorithm. 

\ifx\onecol\undefined
\begin{figure}
	\centering 
	\includegraphics[width=\linewidth]{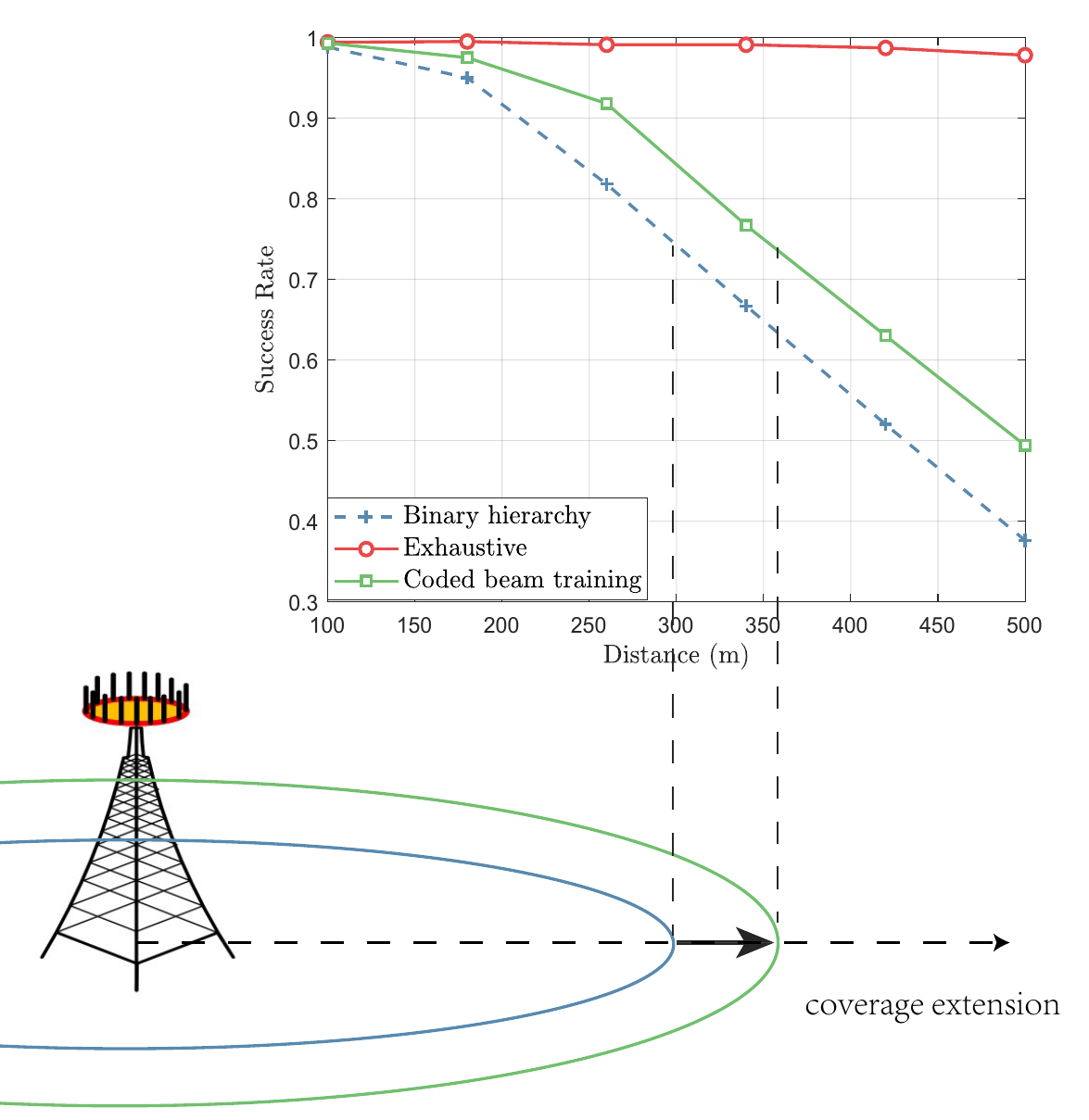}
	\caption{Comparisons of success rate of different beam training methods with different distance, together with the illustration of the extended user coverage. }
	\label{fig:dis}
\end{figure}
\else 
\begin{figure}
	\centering 
	\includegraphics[width=0.6\linewidth]{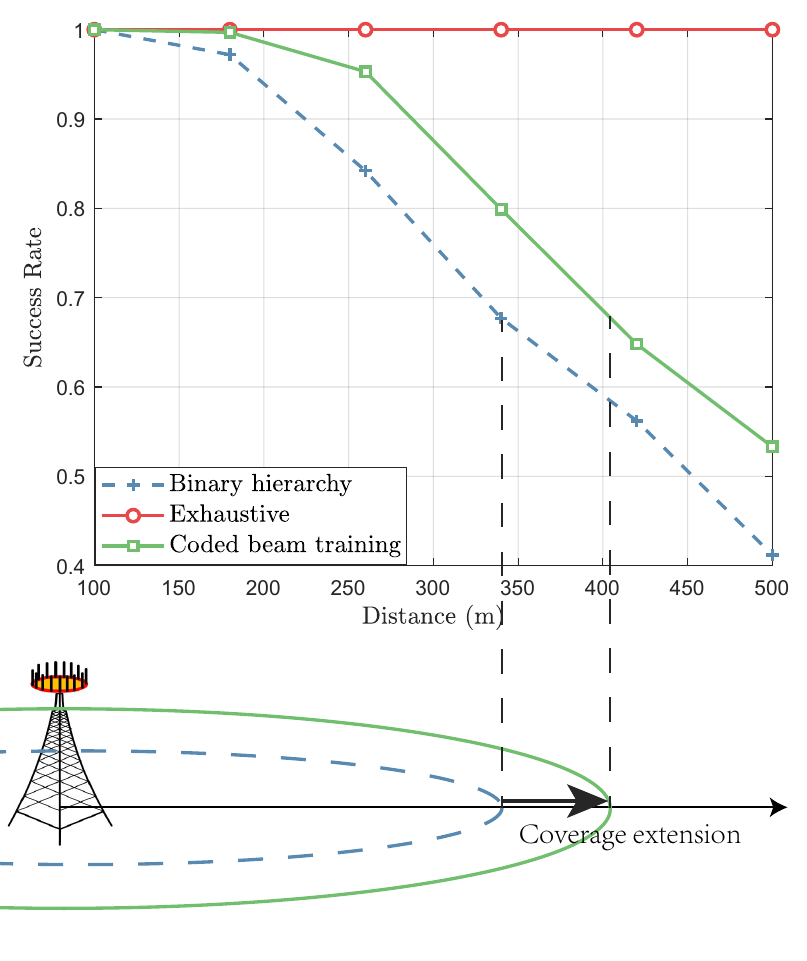}
	\caption{Comparisons of success rate of different beam training methods with different distance, together with the illustration of the extended user coverage. }
	\label{fig:dis}
\end{figure}
\fi

Besides, we present simulation experiments to illustrate the capability of the proposed coded beam training method to extend the coverage area. The simulation results are acquired under carrier frequency $f_c=3.5\, {\rm GHz}$, transmitting power of BS $P_{t}=40\, {\rm dBm}$, bandwith ${\rm BW}=50\, {\rm MHz}$, number of subcarrier $N_{sub}=1024$ and noise power $\sigma^2=-110\, {\rm dBm}$. The large scale fading factor $\gamma=\lambda /(4\pi d_{\rm UE})$, where $\lambda = c/f_c$ is the wave length, and $d_{\rm UE}$ is the distance between BS and UE. As illustrated in Fig.~\ref{fig:dis}, the proposed coded beam training framework can extend coverage area by more than $60 m$ under the same success rate compared with traditional hierarchical beam training. For instance, the proposed coded beam training achieves success rate $0.75$ at $\rm 360\, m$ while traditional hierarchical beam training method can guarantee the same performance only at $\rm 300\, m$. It is promising that the propsed coded beam training method can enable beam training for remote users and thus extend the coverage area.

\section{Conclusions} \label{sec-con}
In this paper, we introduce channel codes into hierarchical beam training to enable reliable implicit CSI acquisition for remote users in future 6G wireless communications. By proving the duality of hierarchical beam training and channel coding, we reveal that the hierarchical beam training problem can be transformed into designing channel codes, which enables the exploitation of the coding gain. We also demonstrate that the decoders need to be modified to fit the beam training problem. Simulation results have verified the effectiveness of the proposed method, which serves as a promising way to achieve reliable coverage of remote users. Future works can be focused on improving the multi-mainlobe beam generation algorithm to produce wide beams with better energy concentration. In addition, diverse channel coding methods can be utilized to further improve the coded beam training performance.

\footnotesize

\bibliographystyle{IEEEtran}
\bibliography{CBT, IEEEabrv}

\begin{thebibliography}{10}
\providecommand{\url}[1]{#1}
\csname url@samestyle\endcsname
\providecommand{\newblock}{\relax}
\providecommand{\bibinfo}[2]{#2}
\providecommand{\BIBentrySTDinterwordspacing}{\spaceskip=0pt\relax}
\providecommand{\BIBentryALTinterwordstretchfactor}{4}
\providecommand{\BIBentryALTinterwordspacing}{\spaceskip=\fontdimen2\font plus
\BIBentryALTinterwordstretchfactor\fontdimen3\font minus
  \fontdimen4\font\relax}
\providecommand{\BIBforeignlanguage}[2]{{%
\expandafter\ifx\csname l@#1\endcsname\relax
\typeout{** WARNING: IEEEtran.bst: No hyphenation pattern has been}%
\typeout{** loaded for the language `#1'. Using the pattern for}%
\typeout{** the default language instead.}%
\else
\language=\csname l@#1\endcsname
\fi
#2}}
\providecommand{\BIBdecl}{\relax}
\BIBdecl

\bibitem{marzetta2010noncooperative}
T.~L. Marzetta, ``Noncooperative cellular wireless with unlimited numbers of
  base station antennas,'' \emph{IEEE Trans. Wireless Commun.}, vol.~9, no.~11,
  pp. 3590--3600, Nov. 2010.

\bibitem{9810144}
X.~Wei, L.~Dai, Y.~Zhao, G.~Yu, and X.~Duan, ``Codebook design and beam
  training for extremely large-scale {RIS}: Far-field or near-field?''
  \emph{China Commun.}, vol.~19, no.~6, pp. 193--204, Jun. 2022.

\bibitem{6G}
T.~S. Rappaport, Y.~Xing, O.~Kanhere, S.~Ju, A.~Madanayake, S.~Mandal,
  A.~Alkhateeb, and G.~C. Trichopoulos, ``Wireless communications and
  applications above 100 {GHz}: Opportunities and challenges for {6G} and
  beyond,'' \emph{IEEE Access}, vol.~7, pp. 78\,729--78\,757, Jun. 2019.

\bibitem{10379539}
Z.~Wang, J.~Zhang, H.~Du, D.~Niyato, S.~Cui, B.~Ai, M.~Debbah, K.~B. Letaief,
  and H.~V. Poor, ``A tutorial on extremely large-scale { MIMO} for {6G}:
  Fundamentals, signal processing, and applications,'' \emph{IEEE Commun. Surv.
  Tutor.}, 2024.

\bibitem{9452036}
X.~Ma, Z.~Gao, F.~Gao, and M.~Di~Renzo, ``Model-driven deep learning based
  channel estimation and feedback for millimeter-wave massive hybrid {MIMO}
  systems,'' \emph{IEEE J. Sel. Area Commun.}, vol.~39, no.~8, pp. 2388--2406,
  Aug. 2021.

\bibitem{9508929}
C.~Han, L.~Yan, and J.~Yuan, ``Hybrid beamforming for {Terahertz} wireless
  communications: Challenges, architectures, and open problems,'' \emph{IEEE
  Wireless Commun.}, vol.~28, no.~4, pp. 198--204, Aug. 2021.

\bibitem{9913211}
Y.~Zhang, X.~Wu, and C.~You, ``Fast near-field beam training for extremely
  large-scale array,'' \emph{IEEE Wireless Commun. Lett.}, vol.~11, no.~12, pp.
  2625--2629, Dec. 2022.

\bibitem{10130629}
W.~Liu, C.~Pan, H.~Ren, F.~Shu, S.~Jin, and J.~Wang, ``Low-overhead beam
  training scheme for extremely large-scale {RIS} in near field,'' \emph{IEEE
  Trans. Commun.}, vol.~71, no.~8, pp. 4924--4940, Aug. 2023.

\bibitem{7438800}
X.~Gao, L.~Dai, Z.~Chen, Z.~Wang, and Z.~Zhang, ``Near-optimal beam selection
  for beamspace mmwave massive mimo systems,'' \emph{IEEE Commun. Lett.},
  vol.~20, no.~5, pp. 1054--1057, May 2016.

\bibitem{exhaustive}
A.~Alkhateeb, G.~Leus, and R.~W. Heath, ``Limited feedback hybrid precoding for
  multi-user millimeter wave systems,'' \emph{IEEE Trans. Wireless Commun.},
  vol.~14, no.~11, pp. 6481--6494, Nov. 2015.

\bibitem{7845674}
J.~Song, J.~Choi, and D.~J. Love, ``Common codebook millimeter wave beam
  design: Designing beams for both sounding and communication with uniform
  planar arrays,'' \emph{IEEE Trans. Commun.}, vol.~65, no.~4, pp. 1859--1872,
  Apr. 2017.

\bibitem{MUexhaus}
X.~Sun, C.~Qi, and G.~Y. Li, ``Beam training and allocation for multiuser
  millimeter wave massive {MIMO} systems,'' \emph{IEEE Trans. Wireless
  Commun.}, vol.~18, no.~2, pp. 1041--1053, Feb. 2019.

\bibitem{tdma}
Z.~Xiao, T.~He, P.~Xia, and X.-G. Xia, ``Hierarchical codebook design for
  beamforming training in millimeter-wave communication,'' \emph{IEEE Trans.
  Wireless Commun.}, vol.~15, no.~5, pp. 3380--3392, May 2016.

\bibitem{9547829}
B.~Ning, Z.~Chen, Z.~Tian, C.~Han, and S.~Li, ``A unified {3D} beam training
  and tracking procedure for {Terahertz} communication,'' \emph{IEEE Trans.
  Wireless Commun.}, vol.~21, no.~4, pp. 2445--2461, Apr. 2022.

\bibitem{10057262}
J.~Wang, W.~Tang, S.~Jin, C.-K. Wen, X.~Li, and X.~Hou, ``Hierarchical
  codebook-based beam training for {RIS}-assisted {mmWave} communication
  systems,'' \emph{IEEE Trans. Commun.}, vol.~71, no.~6, pp. 3650--3662, Jun.
  2023.

\bibitem{10239282}
C.~Wu, C.~You, Y.~Liu, L.~Chen, and S.~Shi, ``Two-stage hierarchical beam
  training for near-field communications,'' \emph{IEEE Trans. Veh. Tech.},
  vol.~73, no.~2, pp. 2032--2044, Feb. 2024.

\bibitem{deact}
J.~Wang, Z.~Lan, C.~woo Pyo, T.~Baykas, C.~sean Sum, M.~Rahman, J.~Gao,
  R.~Funada, F.~Kojima, H.~Harada, and S.~Kato, ``Beam codebook based
  beamforming protocol for multi-{Gbps} millimeter-wave {WPAN} systems,''
  \emph{IEEE J. Sel. Area Commun.}, vol.~27, no.~8, pp. 1390--1399, Oct. 2009.

\bibitem{channel}
A.~Alkhateeb, O.~El~Ayach, G.~Leus, and R.~W. Heath, ``Channel estimation and
  hybrid precoding for millimeter wave cellular systems,'' \emph{IEEE J. Sel.
  Top. Signal Process.}, vol.~8, no.~5, pp. 831--846, Oct. 2014.

\bibitem{sum}
L.~Chen, Y.~Yang, X.~Chen, and W.~Wang, ``Multi-stage beamforming codebook for
  {60GHz WPAN},'' in \emph{2011 6th International ICST Conference on
  Communications and Networking in China (CHINACOM)}, 2011, pp. 361--365.

\bibitem{8408778}
Z.~Xiao, H.~Dong, L.~Bai, P.~Xia, and X.-G. Xia, ``Enhanced channel estimation
  and codebook design for millimeter-wave communication,'' \emph{IEEE Trans.
  Veh. Tech.}, vol.~67, no.~10, pp. 9393--9405, Oct. 2018.

\bibitem{Hybrid}
E.~Zhang and C.~Huang, ``On achieving optimal rate of digital precoder by
  {RF}-baseband codesign for {MIMO} systems,'' in \emph{Proc. 2014 IEEE 80th
  Vehicular Technology Conference (IEEE VTC'14 Fall)}, 2014, pp. 1--5.

\bibitem{9411813}
S.~Lyu, Z.~Wang, Z.~Gao, H.~He, and L.~Hanzo, ``Lattice-based mmwave hybrid
  beamforming,'' \emph{IEEE Trans. Commun.}, vol.~69, no.~7, pp. 4907--4920,
  Jul. 2021.

\bibitem{9618146}
W.~Xu, L.~Gan, and C.~Huang, ``A robust deep learning-based beamforming design
  for ris-assisted multiuser {MISO} communications with practical
  constraints,'' \emph{IEEE Trans. Cogn. Commun. Netw.}, vol.~8, no.~2, pp.
  694--706, Jun. 2022.

\bibitem{10294206}
G.~Sun, W.~Yan, W.~Hao, C.~Huang, and C.~Yuen, ``Beamforming design for the
  distributed {RIS}s-aided thz communications with double-layer true time
  delays,'' \emph{IEEE Trans. Veh. Tech.}, pp. 1--15, 2023.

\bibitem{beamgene}
K.~Chen, C.~Qi, and G.~Y. Li, ``Two-step codeword design for millimeter wave
  massive {MIMO} systems with quantized phase shifters,'' \emph{IEEE Trans.
  Signal Process.}, vol.~68, pp. 170--180, Jan. 2020.

\bibitem{10365224}
Y.~Lu, Z.~Zhang, and L.~Dai, ``Hierarchical beam training for extremely
  large-scale {MIMO}: From far-field to near-field,'' \emph{IEEE Trans.
  Commun.}, 2024.

\bibitem{shannon1948mathematical}
C.~E. Shannon, ``A mathematical theory of communication,'' \emph{Bell Syst.
  Tech. J}, vol.~27, no.~3, pp. 379--423, Jul. 1948.

\bibitem{7947209}
S.~Noh, M.~D. Zoltowski, and D.~J. Love, ``Multi-resolution codebook and
  adaptive beamforming sequence design for millimeter wave beam alignment,''
  \emph{IEEE Trans. Wireless Commun.}, vol.~16, no.~9, pp. 5689--5701, 2017.

\bibitem{10033092}
B.~Ning, T.~Wang, C.~Huang, Y.~Zhang, and Z.~Chen, ``Wide-beam designs for
  {T}erahertz massive {MIMO}: {SCA-ATP} and {S-SARV},'' \emph{IEEE Internet
  Things J.}, vol.~10, no.~12, pp. 10\,857--10\,869, Jun. 2023.

\bibitem{conv}
P.~Elias, ``Coding for noisy channels,'' \emph{IRE Convention Record}, vol.~7,
  pp. 37--47, May 1955.

\bibitem{gs}
O.~Bucci, G.~Franceschetti, G.~Mazzarella, and G.~Panariello, ``Intersecetion
  approach to array pattern synthesis,'' \emph{IEEE Internet Things J.}, vol.
  137, no.~6, pp. 349--357, Apr. 1990.

\bibitem{DAVIDFORNEY1974222}
G.~{David Forney}, ``Convolutional codes {II}. maximum-likelihood decoding,''
  \emph{Information and Control}, vol.~25, no.~3, pp. 222--266, Jun. 1974.

\end{thebibliography}

\normalsize



\end{document}